\documentclass[acmsmall,authorversion,nonacm]{acmart}

\usepackage[english]{babel}
\usepackage{blindtext}

\usepackage{subcaption}
\usepackage{makecell}
\usepackage{wrapfig}

\usepackage{tikz}
\usetikzlibrary{positioning}
\usepackage{varwidth}

\usepackage{amsmath}

\newcommand{\note}[1]{}

\usepackage{xspace}
\newcommand*{\eg}{e.g.,~}

\newcommand*{\ie}{\textit{i.e.,}~}

\newcommand{\squishlist}{
 \begin{list}{${\bullet}$}
  { \setlength{\itemsep}{0pt}
     \setlength{\parsep}{1pt}
     \setlength{\topsep}{1pt}
     \setlength{\partopsep}{0pt}
     \setlength{\leftmargin}{1em}
     \setlength{\labelwidth}{0.5em}
     \setlength{\labelsep}{0.5em} } }

\newcommand{\squishend}{
  \end{list}  }

\AtBeginDocument{%
  }

\begin{document}

\title[The Multifractal IP Address Structure]{The Multifractal IP Address Structure: \\ Physical Explanation and Implications}

\author{Chris Misa}
\affiliation{\institution{University of Oregon}
  \city{}
  \country{USA}
}

\author{Ram Durairajan}
\affiliation{\institution{University of Oregon}
  \city{}
  \country{USA}
}

\author{Arpit Gupta}
\affiliation{\institution{UCSB}
  \city{}
  \country{USA}
}

\author{Reza Rejaie}
\affiliation{\institution{University of Oregon}
  \city{}
  \country{USA}
}

\author{Walter Willinger}
\affiliation{\institution{NIKSUN, Inc.}
  \city{}
  \country{USA}
}

\authorsaddresses{}

\begin{abstract}
The structure of IP addresses observed in Internet traffic plays a critical role for a wide range of networking problems of current interest.
For example, modern network telemetry systems that take advantage of existing data plane technologies for line rate traffic monitoring and processing cannot afford to waste precious data plane resources on traffic that comes from ``uninteresting” regions of the IP address space.
However, there is currently no well-established structural model or analysis toolbox that enables a first-principles approach to the specific problem of identifying ``uninteresting'' regions of the address space or the myriad of other networking problems that prominently feature IP addresses.

To address this key missing piece, we present in this paper a first-of-its-kind empirically validated physical explanation for why the observed IP address structure in measured Internet traffic is multifractal in nature. Our root cause analysis overcomes key limitations of mostly forgotten findings from $\sim$20 years ago and demonstrates that the Internet processes and mechanisms responsible for how IP addresses are allocated, assigned, and used in today’s Internet are consistent with and well modeled by a class of evocative mathematical models called conservative cascades. We complement this root cause analysis with the development of an improved toolbox that is tailor-made for analyzing finite and discrete sets of IP addresses and includes statistical estimators that engender high confidence in the inferences they produce. We illustrate the use of this toolbox in the context of a novel address structure anomaly detection method we designed and conclude with a discussion of a range of challenging open networking problems that are motivated or inspired by our findings.

\end{abstract}

\maketitle

\section{Introduction}

The spatial structure of observed IP addresses in measured packet traces of Internet traffic is a manifestation of how IP addresses are allocated, assigned, and used in the Internet.\footnote{Note that ``spatial'' here refers to the ``address space'' which is the set of all possible IPv4 or IPv6 addresses and which is distinct from other notions of spatial structure (\eg~\cite{avin2020complexity}).} Understanding this structure means providing quantitative information on the patterns formed by the IP addresses responsible for observed traffic and is critical for a wide range of current and future Internet traffic engineering and network management problems.
For example, systems that monitor traffic by computing per-address summary statistics directly in limited-resource in-network processors (\eg switch ASCIs, IPUs/DPUs) must leverage address structure to maximize efficiency of limited memory resources by avoiding ``uninteresting'' address space regions.
Likewise, security systems that block or limit addresses using publicly-established ``reputation'' based on past behavior must be able to assess in real time when newly observed addresses do or do not fit into the address structure commonly observed at a particular vantage point.
In the case of future AI systems intended for automating different network management tasks, it will be important to provide quantitative metrics to leverage critical correlations embedded in address structure (\eg as intuitively identified by human operators with year of experience working with IP addresses) while avoiding the danger of overfitting on specific addresses observed in training data.

Supporting these and other example tasks, as well as potential future innovations, implies that a useful understanding and characterization of observed IP address structure must meet the following three requirements.

\squishlist
\item 
Provides quantitative methods and metrics that succinctly summarize local structure of populated address space regions and/or global structure of encountered sets of addresses. 

\item 
Establishes high confidence in the physical meaning of the quantitative methods and metrics by identifying, empirically validating, and accurately modeling the mechanisms responsible for the provided physical explanation. 

\item
Engenders high confidence in the statistical methods used to infer the relevant model parameters and derived metrics from real-world data given in the form of sets of observed IP addresses.
\squishend

Although characterizing the structure of observed IP addresses has received some attention in the past, to the best of our knowledge, no prior research efforts have provided satisfactory responses to all the above three requirements.
The most prominent line of this prior work is from the early 2000s~\cite{kohler2002observed,barford2006toward} and presented preliminary empirical evidence that this structure exhibits {\em multifractal scaling} behavior and can therefore be modeled by means of mathematical objects called multifractals. However, this and similar works around that time limited their efforts to applying commonly-used but poorly-understood multifractal analysis techniques and steered clear of pursuing a physical explanation or root cause analysis of the observed behavior; that is, identifying the real-world Internet processes and mechanisms that cause the structure of observed IP addresses to be multifractal in nature. The importance of such a physical explanation is succinctly summed up by the authors of~\cite{kohler2002observed}: \textit{``without a convincing description of how [multifractal] address structure arises, the results of the explorations [in~\cite{kohler2002observed}] must be considered preliminary.”}

Our work revisits the question of whether or not observed IP addresses in measured Internet packet traces exhibit multifractal scaling behavior. However, compared to prior work, we approach the question from a different angle and present several novel contributions that we detail below. Together, these contributions allow us to not only answer the question in the affirmative but also establish the root cause for the observed multifractal structure and its practical relevance.

\squishlist
\item 
Rather than proceeding directly with statistical analysis, in \S~\ref{sec:physical-explanation} we first develop a direct association between the process of IP allocation and assignment (which ultimately dictates which addresses are observed in the Internet) and a mathematical construct known as {\em conservative cascade model}, and support this association with empirical evidence from real-world address space allocation policies.

\item
We demonstrate in \S~\ref{sec:generative} that the finite and discrete nature of IP spaces has important ramifications and prevents the ``out-of-the-box" application of existing generative methods that have been specifically developed for synthesizing mathematical multifractals using conservative cascade models. We show how these techniques that assume an infinite, continuous underlying space can be adapted to generate finite sets of discrete IP addresses that closely mimic the properties of real-world observed sets of addresses.

\item
We consider in \S~\ref{sec:evidence} a known approach to inferring multifractal scaling from real-world data referred to as the method-of-moments and improve it to account for the practical limitations that finite and discrete IP spaces pose for methods originally defined for continuous and infinite spaces. In particular, we adapt an existing statistical estimator to overcome these limitations and use it to demonstrate and quantify the existence of multifractal scaling in a wide range of real-world observed IP address data.

\item
We detail in \S~\ref{sec:implications} an illustrative example of a possible use case for quantifying structural properties of sets of IP addresses by developing a novel address structure anomaly detection method based on our improved statistical estimator.
This method formally quantifies the common knowledge among network operators that observing IP addresses from new, less-common regions is often correlated with security risks and provides a tool for assessing the perils associated with a newly observed IP address that goes beyond the use of simple IP reputation lists.
\squishend

We conclude in \S~\ref{sec:discussion} with a discussion of a wide range of potential impacts of firmly establishing multifractal scaling as a predominant property and of the spatial structure of observed IP addresses and a new invariant of measured Internet traffic.
The analysis code, scripts, and some data sets will be made available upon publication.

\section{Background and Motivation}
\label{sec:background}

\subsection{An echo from the past}
\label{ssec:pictorial-evidence}

Preliminary findings about an apparent multifractal structure of observed IP addresses in measured network traffic were originally reported in~\cite{kohler2002observed,kohler2006observed,barford2006toward} and rely on traffic traces that have been collected some 20-25 years ago. To address the dated nature of these traces, we first set out to assemble our own repository of more recent Internet traffic traces. A detailed description of this repository can be found in Appendix~\ref{appx:traceRepo}, with Table~\ref{table:list_of_datasets} listing all the traces and associated metadata. 

We then use one of these more recent traces, CAIDA-dir-A, and provide pictorial evidence that the observations made in~\cite{kohler2002observed,kohler2006observed} still hold some 20 years later.
Using the CAIDA-dir-A trace, we form a dataset CAIDA500k of IPv4 addresses that consists of the first $500,000$ unique source IP addresses observed in CAIDA-dir-A.
As a point of comparison, we also generate a same-sized synthetic dataset Uniform500k by casting 500k 32-bit integers as IP addresses and drawing them from the uniform distribution on the unit interval.

Following the adage ``a picture is worth a thousand words,'' we visualize the structure of IPv4 addresses by mapping each address to a subset of the interval $[0, 1) \subset \mathcal{R}$.
This mapping is formed by associating IPv4 address prefixes (\eg specified in CIDR notation) to iterative binary subdivisions of $[0, 1)$.
For example 0.0.0.0/1 corresponds to $[0, 1/2)$, 128.0.0.0/1 corresponds to $[1/2, 1)$, 0.0.0.0/2 corresponds to $[0, 1/4)$ and so forth.
This mapping offers a compact, visual way to reason about IPv4 addresses and address prefixes while also facilitating application of continuous mathematical methods defined on $\mathcal{R}$.

\begin{figure}[htb!]
    \vspace{-0.3cm}
    \centering
    \begin{subfigure}[t]{0.445\linewidth}
        \centering
        \includegraphics[width=1.0\linewidth]{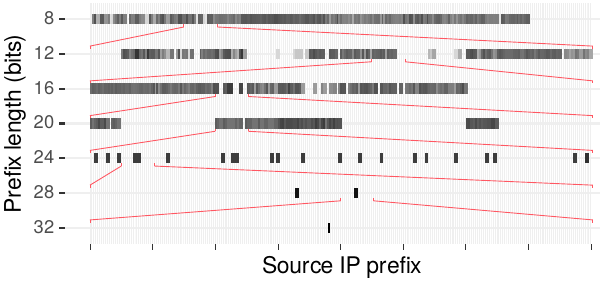} %
        \caption{\label{1Dfigure:compare:caida}CAIDA500k}
    \end{subfigure}~
    \begin{subfigure}[t]{0.09\linewidth}
        \vspace{-2.5cm}
        \includegraphics[width=1.0\linewidth]{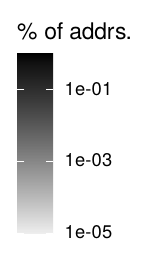}
    \end{subfigure}~
    \begin{subfigure}[t]{0.445\linewidth}
        \centering
        \includegraphics[width=1.0\linewidth]{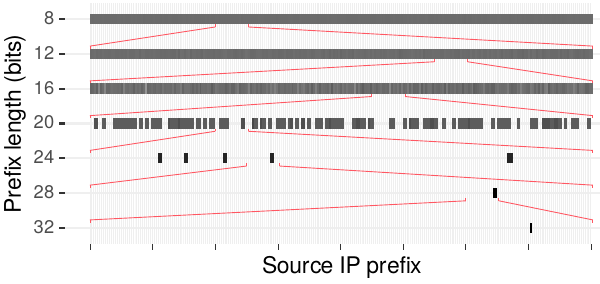}
        \caption{\label{1Dfigure:compare:uniform}Uniform500k.}
    \end{subfigure}
    \vspace{-0.4cm}
    \caption{1D Pictorial evidence of multifractal vs not-multifracatal IP address structure, using the dataset CAIDA500k (left) and Uniform500k (right), respectively.} 
    \label{1Dfigure:compare}
    \vspace{-0.3cm}
\end{figure}

Next, building on the intuitive notion that multifractal structure manifests as a pronounced ``cluster-within-cluster" property, we draw a series of ``zoom-in" steps that reveal how clusters of observed IP addresses at a given scale (\eg /16 subnet granularity) appear when viewed under a microscope (\eg /20 or finer subnet granularity). Figure~\ref{1Dfigure:compare} illustrates six steps of this ``zoom-in" process for the real-world dataset CAIDA500k (Figure~\ref{1Dfigure:compare:caida}) and the synthetic dataset Uniform500k (Figure~\ref{1Dfigure:compare:uniform}), respectively.

In the case of the CAIDA500k dataset (Figure~\ref{1Dfigure:compare:caida}),
the zoom-in’s for the intermediate stages look qualitatively similar (with the exception of artifacts in the top row and bottom two rows) and exhibit visually apparent cluster-within-cluster behavior with no ``typical'' cluster size (\eg in terms of their width on the x-axis of their number of addresses shown by shades of gray), providing pictorial evidence consistent with a multifractal address structure.
In the case of the Uniform500k dataset (Figure~\ref{1Dfigure:compare:uniform}),
we also observe qualitatively similar-looking zoom-ins across the different rows, but the appearance of the zoom-ins is noticeably less ``intermittent''---the cluster-within-cluster behavior is more regular, allows for discerning a ``typical” cluster size for each row, providing evidence consistent with monofractal structure.

The visual evidence in Figure~\ref{1Dfigure:compare:caida} suggests that the observed IP addresses in measured Internet traffic are distributed across the IPv4 space in a highly irregular or intermittent manner, symptomatic for an address structure that exhibits the hallmarks of mathematical objects called {\bf multifractal measures} (or multifractals, for short)~\cite{evertsz1992multifractal,salat2017multifractal}. 
We summarize two key required concepts here only at an intuitive level and leave an in-depth introduction of the relevant mathematical concepts and constructions to Appendix~\ref{sec:appx-mathBackground}.

First, {\bf multifractal analysis} at its core seeks to characterize measures (\eg measuring the number of observed addresses per IP address prefix) by defining and estimating metrics that effectively capture and summarize how sets at one scale (\eg prefixes at one prefix length) and their associated mass fragment into subsets and masses at the next finer scale for increasingly finer scales. As such, these metrics are meant to describe the ``scaling behavior" or ``clustering structure" of measures as they assign mass to increasingly smaller sets.
The two most common such metrics developed in the statistics community are referred to as the {\bf multifractal spectrum}, written $f(\alpha)$, and the {\bf structure function}, written $\tau(q)$.
When applied to a multifractal measure (\ie complex scaling behavior), $f(\alpha)$ is defined for a range of $\alpha$-values on $\mathcal{R}$ and takes on an bell-like shape whereas $\tau(q)$ is typically defined for a range of real-valued $q$-values and takes on a concave-down curved shape.
When applied to a monofractal measure (\ie simple scaling behavior), $f(\alpha)$ is degenerate (\ie only defined at a single $\alpha$) and $\tau(q)$ is linear in $q$.
Although these two metrics are directly connected in theory (by the Legendre transform), in practice one cannot effectively convert an estimate of one into an estimate of the other.

Second, a {\bf conservative cascade} is an iterative process that splits a set into subsets and simultaneously divides a measure of mass on that set between the subsets based on rules defined in terms of a generator distribution $W$.
Under certain conditions, conservative cascades produce multifractal measures whose entire scaling behavior is determined exclusively by the generator $W$.
This connection enables identification of a physical explanation for the observed multifractal scaling behavior in real-world data by inferring the conservative cascade that presumably generated this data in the first place. 
Moreover, recent work in the statistics community develops powerful tools (\eg~\cite{ossiander2000statistical}) for estimating the structure functions of conservative cascades based on observation of a single cascade realization.

\subsection{Reviving the echo: contributions and roadmap}
\label{ssec:limitations}

Three key limitations attenuated the potentially wider impact of the original efforts reported in~\cite{kohler2002observed,kohler2006observed} to characterize the structure of observed IP addresses.

\noindent
{\bf Limitation 1: Lack of physical explanation for multifractal IP address structure.}
A few prior works make informal assessments that certain facts about how IPv4 addresses are allocated could potentially explain why they are observed to exhibit multifractal scaling behavior~\cite{kohler2002observed,barford2006toward}.
However, these studies provided neither a theoretical foundation nor any empirical evidence to support these claims leading to inconclusive conclusions.

\noindent
{\bf Limitation 2: Lack of confidence in statistical analysis method.}
Prior works including~\cite{kohler2002observed,barford2006toward}
leverage histogram-based methods (the most common and best-understood methods at their time of publication) to directly estimate the multifractal spectrum of limited sets of observed IPv4 addresses.
As suggested above (and in~\cite{evertsz1992multifractal}), the well-documented limitations of the histogram method lead to inconclusive evidence for multifractal scaling in these works (in particular the produced multifractal spectra are noisy and jagged with no easily-discernible width).
A key focus in the statistical community in the years since these prior works has been to develop more robust methods to determine more conclusive evidence for the presence of multifractal scaling in real-world data (see for example~\cite{ossiander2000statistical,gilbert-etal:2003,wendt-abry:2007}).

\noindent
{\bf Limitation 3: Lack of usecase pull.}
Previous works do not adequately explore the potential impact of the multifractal structure of observed IP addressed on critical challenges in networking research and system design.
Moreover, they lack compelling evidence to demonstrate how this structure could influence these areas.
This gap limits understanding of the significance of multifractal structure and impedes the broader adoption of associated theories and techniques in both operational and research contexts.

\smallskip
\noindent \textbf{\em Contributions and roadmap.} To revive and amplify the limited initial explorations of prior works and to establish a strong foundation for understanding multifractal scaling of observed IP addresses, our work addresses each of these limitations in turn.
First, in \S~\ref{sec:physical-explanation} we use data from IANA and bulk WHOIS records from each of the RIRs to establish that the process of IP address allocation at different levels is consistent with a non-trivial conservative cascade model thereby addressing {\bf Limitation 1}.
Second, in \S~\ref{sec:generative} we leverage the conservative cascade model to build a generator of realistic IP addresses by solving several challenges arising from the finite discreteness of IP addresses.
Third, in \S~\ref{sec:evidence} we build on modern method-of-moments approaches for determining the presence of multifractal scaling and estimating the ``structure function'' ($\tau(q)$ as discussed in \S~\ref{ssec:multifracatal-analysis}) to address {\bf Limitation 2}.
Finally, in \S~\ref{sec:implications} we address {\bf Limitation 3} by illustrating how our approach for measuring IP address structure can be leveraged to detect the arrival of anomalous addresses in an efficient streaming anomaly detection algorithm.
Section \S~\ref{sec:discussion} further addresses {\bf Limitation 3} through qualitative discussion of a wider range of networking problems potentially impacted by multifractal scaling of observed IP addresses.

\section{Physical Explanation and Validation}
\label{sec:physical-explanation}

\begin{figure}[!htb]
    \vspace{-0.3cm}
    \centering
    \includegraphics[width=0.7\columnwidth]{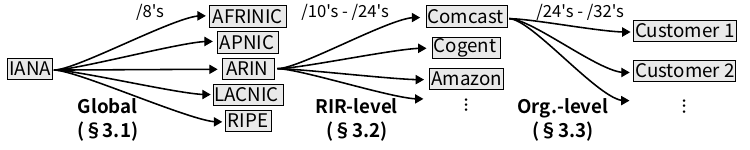}
    \vspace{-0.3cm}
    \caption{High-level ``cascade'' process of Internet address allocations.}
    \label{fig:phyExplOverview}
    \vspace{-0.3cm}
\end{figure}

In this section, we hypothesize that the visual evidence of multifractal scaling in the structure of observed IP addresses, suggested in Figure~\ref{1Dfigure:compare:caida}, is a direct result of address allocation policies and investigate this hypothesis through empirical analysis of publicly-available address allocation datasets.
To this end, we are the first to explicitly show the presence of strong hierarchic clustering behavior in address space allocations {\em beyond} the classic ``three-level story'' and suggest the address allocation process can be well-modeled by a conservative cascade.
Here, the classic ``three-level story'' refers to the process depicted in Figure~\ref{fig:phyExplOverview}: IANA makes ``global'' allocations to the Regional Internet Registries (RIRs), RIRs make ``first-level'' allocations to particular service-provider-type organizations (\eg ISPs, CSPs) sometimes known as Local Internet Registries (LIRs), and service-providers or LIRs make allocations or assignments to their customers.\footnote{Note that {\em allocations} are normally understood to refer to intermediate decisions in this process (\ie address ranges to be further divided by a down-stream entity) whereas {\em assignments} are terminal decisions (\ie to be used for the infrastructure of a particular end-user entity)~\cite{ripeAllocVsAssign,otherAllocVsAssign}.} 
In the following, we demonstrate that at each level of this story, more detailed cascade (sub-)processes are also evident in how particular entities organize their address spaces, revealing additional iterations in a non-trivial cascade construction.
We perform all analysis for both IPv4 and IPv6 and provide supporting evidence and additional details for some claims in Appendix~\S~\ref{sec:appdx-physical-explanation}.

\subsection{IANA: Allocation policy at coarse granularity} 
\label{ssec:iana}

The top-level authority responsible for the total ``mass'' of all IP addresses is the Internet Assigned Numbers Authority (IANA).
IANA divides global address spaces by coordinating and publishing allocations of address ranges for management by particular RIRs and for particular technical usecases (\eg multicast protocols, local address space).

\noindent
{\bf Global organization of IPv4.}
For the IPv4 address space, IANA publishes an authoritative list of how each possible top-level /8 prefix is allocated~\cite{ianaipv4} known as the ``IPv4 Address Space Registry''. 
After allocation of the last free /8 block in 2011~\cite{ipv4exhaustion}, the %
registry now covers the entire IPv4 address space.

The fact that all IANA IPv4 allocations are made at /8 granularity may appear to imply a single iteration of the cascade process (\ie IANA-level $\rightarrow$ RIR-level).
However, closer inspection of how /8's are allocated to particular RIRs reveals a significant degree of clustering along boundaries implied by shorter prefix lengths (\eg the /1, /2, and /3 clusters leftover from class-{\em full} network allocation or induced by reserved spaces) indicating several additional levels of hierarchy in how IANA organizes /8 allocations.

\begin{figure}[!htb]
    \vspace{-0.3cm}
    \centering
    \includegraphics[width=0.68\columnwidth]{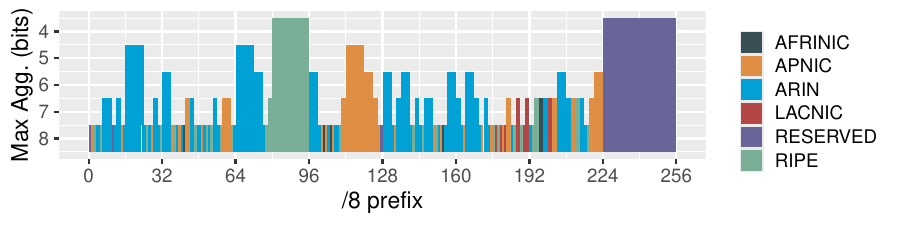}
    \vspace{-0.3cm}
    \caption{Global view of IANA IPv4 Address Space Registry showing which regions of the space are allocated to several primary RIRs and the maximum possible aggregation of each contiguous per-RIR block.}
    \label{fig:iana_maxAgg}
    \vspace{-0.3cm}
\end{figure}

Figure~\ref{fig:iana_maxAgg} shows which RIR is associated with each /8 block allocated by IANA\footnote{We use the WHOIS field of each block instead of the Designation field here since several /8 prefixes are still listed as allocated to individual companies or organizations (\eg the US Department of Defense) while delegating WHOIS record management to a parent RIR.} (x-axis) and how much aggregation is possible in each contiguous cluster of blocks allocated to the same RIR (y-axis).
Note that smaller Max Aggregation Lengths indicate shorter prefixes with more aggregation.
The largest possible aggregation (outside of reserved space) is RIPE's cluster around 80.0.0.0/4 and both ARIN and APNIC also have large /5 clusters.
Our analysis also shows that this type of allocation has clear clustering along the historic class-full network boundaries and that several clusters were allocated sequentially over several years indicating that IANA likely pre-allocated regions for particular RIRs (at shorter than /8 granularity, see Appendix~\ref{ssec:appx-historicalv4}).
The existence of this intra-IANA structure at, \eg /4 granularity implies {\em at least one} additional cascade iteration within IANA's IPv4 allocations (\ie IANA $\rightarrow$ per-RIR clusters $\rightarrow$ /8 allocations).

\begin{figure}[!htb]
\vspace{-0.3cm}
    \centering
    \includegraphics[width=0.65\textwidth]{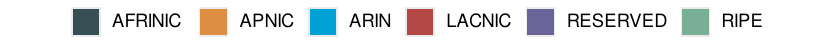}
    \begin{subfigure}{0.48\textwidth}
        \centering
        \includegraphics[width=\columnwidth]{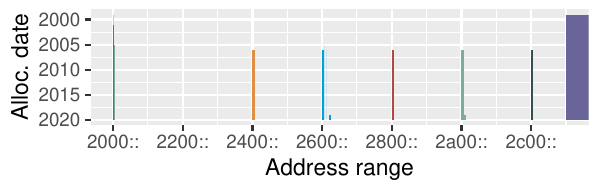}
        \vspace{-0.5cm}
        \caption{Entire unicast space.}
        \vspace{-0.3cm}
    \end{subfigure}~
    \begin{subfigure}{0.48\textwidth}
        \centering
        \includegraphics[width=\columnwidth]{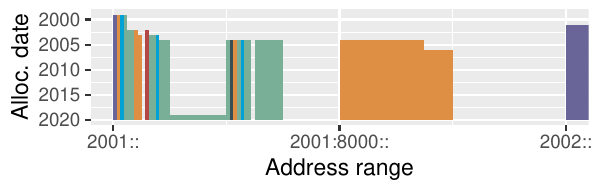}
        \vspace{-0.5cm}
        \caption{Detail of early allocation region.}
        \vspace{-0.3cm}
    \end{subfigure}
    \caption{Timeline of IANA allocations for the non-reserved regions of the global unicast IPv6 space 2000::/3 showing how particular RIRs receive allocations in well-defined clusters.}
    \label{fig:iana_v6Timeline}
    \vspace{-0.3cm}
\end{figure}

\noindent
{\bf Global organization of IPv6}
Similar to IPv4, IANA manages global allocations in the IPv6 address space and maintains, in particular, a registry of which global unicast IPv6 address ranges have been allocated to each RIR~\cite{ianaipv6}.\footnote{Currently IANA only allocates addresses from 2000::/3 and other ranges are reserved for different purposes or future use.}
Figure~\ref{fig:iana_v6Timeline} shows which particular blocks have been allocated thus far (x-axis) along with their allocation dates (y-axis) for the non-reserved region of the global unicast space 2000::/3.

Besides the obviously sparse allocation compared to IPv4, this plot reflects a clear change in IPv6 allocation policy at IANA in 2006.
Prior to 2006, IANA allocated smaller /23 blocks from 2001::/16 following a policy with no clear RIR-level distinctions.
In 2006, following a policy change to allocate larger IPv6 blocjs~\cite{spacebetween}, IANA allocated one large /12 block to each RIR.
As Figure~\ref{fig:iana_v6Timeline} clearly illustrates, these allocations are spaced out such that each RIR has a much larger contiguous /7 block surrounding it indicating this policy change also implemented a {\em global organization} of the IPv6 unicast space along RIR boundaries.
Similar to IPv4, this yields strong evidence for {\em at least one} additional cascade iteration reflecting clear per-RIR clusters before /12 allocations.

\noindent
\textbf{\em Summary.} Clear prefix-level clustering of allocations made to particular RIRs indicates at least one additional level of hierarchy in the internal process IANA implements to divide the ``mass'' of its allocations across the global IPv4 and IPv6 address spaces.

\subsection{The RIRs: Allocation policy at medium granularity}
\label{ssec:arin-allocs}

We next investigate how the allocation policies of particular RIRs contribute to the conservative cascade model of IP address allocation.
RIRs re-allocate and assign sub-regions of the address space allocated to them by IANA and keep track of these allocations and assignments in a global WHOIS database as ``network'' records.
A network record specifies which particular organization is responsible for a particular range of addresses.
Network records document not only direct allocations from RIRs to governments or large service providers, but also sub-allocations and assignments that iteratively divide the space into smaller ranges associated with particular end-users.
To empirically measure how many iterations this well-known cascade process introduces (\ie beyond the simple RIR $\rightarrow$ organization iteration), we analyze snapshots of all WHOIS records from four RIRs known as ``bulk WHOIS'' datasets.

\begin{wraptable}{R}{0.5\textwidth}
    \vspace{-0.3cm}
    \centering
    \scalebox{0.9}{
    \begin{tabular}{l|r|r}
    {\bf RIR} & {\bf IPv4 prefixes} & {\bf IPv6 prefixes}\\
    \hline
    AFRINIC & 125k & 32k \\
    APNIC & 1.256M & 107k \\
    ARIN & 3.147M & 285k \\
    RIPE & 4.160M & 890k \\
    \end{tabular}}
    \caption{Number of distinct prefixes in our bulk WHOIS dataset per-RIR and IP protocol version.}
    \label{tab:whoisPrefixes}
    \vspace{-0.5cm}
\end{wraptable}

\noindent
{\bf Dataset and methodology.}
We collect bulk WHOIS data from four RIRs\footnote{Similar to~\cite{hsu2023fiat}, we attempted to obtain the LACNIC bulk WHOIS dataset, but failed due to administrative impasse.} and extract $\sim$8.7M IPv4 and $\sim$1.3M IPv6 network records.
We filter based on the IANA address ranges allocated to each RIR to remove irrelevant records (\eg placeholders for the global IP address space or for ranges not managed by each particular RIR) that could otherwise inflate our measurement of depth of hierarchy.
Because many RIRs specify IPv4 networks as a range of 32-bit integers (rather than CIDR-based prefixes), we infer the prefix(es) referred to by these ranges (\eg the range 0.0.0.0 to 0.255.255.255 refers to 0.0.0.0/8). In cases where a range does not map to a single prefix, we generate a list of the largest possible prefixes that fit in this range (\eg the range 0.3.0.0 to 0.5.255.255 refers to 0.3.0.0/16 and 0.4.0.0/15).
Only $\sim$0.7\% of records from all RIRs required such treatment.
Table~\ref{tab:whoisPrefixes} shows how many total prefixes we collected through this method for each RIR and IP protocol version.

\noindent
{\bf The shape of RIR-level allocations and assignments.}
We quantify the number of explicit cascade iterations in the RIR-level address allocation process by constructing the {\em prefix-inclusion tree}.
Each WHOIS network record is a node in this tree and if a record is a direct sub-prefix of another record (\ie with no other sub-prefixes in between), then it is that record's child.
We next summarize the ``width'' and ``depth'' of this tree which characterize the amount of mass-splitting at each iteration and number of iterations respectively.

\begin{figure}[!htb]
    \vspace{-0.3cm}
    \centering
    \includegraphics[width=0.45\linewidth]{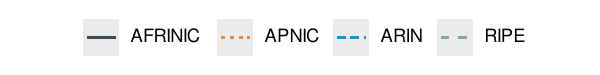}
    
    \begin{minipage}{0.48\linewidth}
        \centering
        \begin{subfigure}[t]{0.48\linewidth}
            \includegraphics[width=1.0\columnwidth]{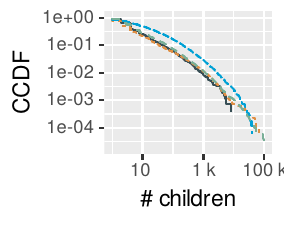}
            \vspace{-0.5cm}
            \caption{IPv4}
            \label{fig:whoisNumChildren:v4}
            \vspace{-0.2cm}
        \end{subfigure}~\hspace{0.03\linewidth}
        \begin{subfigure}[t]{0.48\linewidth}
            \includegraphics[width=1.0\columnwidth]{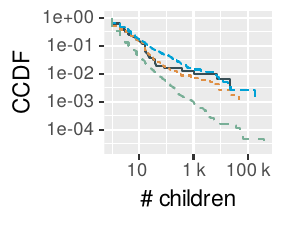}
            \vspace{-0.5cm}
            \caption{IPv6}
            \label{fig:whoisNumChildren:v6}
            \vspace{-0.2cm}
        \end{subfigure}
        \caption{Distribution of number of children over all ``parent'' records.}
        \label{fig:whoisNumChildren}
    \end{minipage}\hfill
    \begin{minipage}{0.48\linewidth}
        \centering
        \begin{subfigure}[t]{0.48\linewidth}
            \includegraphics[width=1.0\columnwidth]{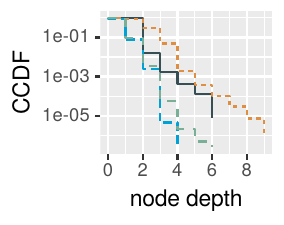}
            \vspace{-0.5cm}
            \caption{IPv4}
            \vspace{-0.2cm}
        \end{subfigure}~\hspace{0.03\linewidth}
        \begin{subfigure}[t]{0.48\linewidth}
            \includegraphics[width=1.0\columnwidth]{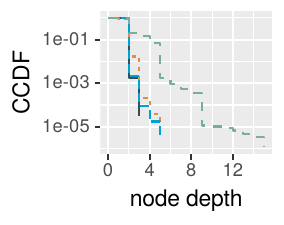}
            \vspace{-0.5cm}
            \caption{IPv6}
            \vspace{-0.2cm}
        \end{subfigure}
        \caption{Distribution of depth in prefix tree over all records.}
        \label{fig:whoisDepth}
    \end{minipage}
    \vspace{-0.2cm}
\end{figure}

To measure width of the prefix-inclusion tree, we examine how ``parent'' records---defined as records that contained one or more child records---split their address-range ``mass'' among their children.
Overall, our dataset has $\sim$89k IPv4 parent records and $\sim$45k IPv6 parent records.
Figure~\ref{fig:whoisNumChildren} shows that the distribution of node degree (\ie the number of child records) over all parent records is pronouncedly long-tailed with median of 4 (1) going up to 95th percentile at 227 (10) for IPv4 (IPv6).
Through manual inspection of samples of 50 parent records above and below the median, we confirmed intuitions about how different numbers of children reflect different types of organizations (\eg parents with large numbers of children tend to correspond to large-scale ISPs or hosting providers).

To measure depth of the prefix-inclusion tree, we apply the standard definition of tree depth (\ie a node's depth is the number of edges from that node to the root of the tree---in this case the top-most network record).
Figure~\ref{fig:whoisDepth} shows that the prefix-inclusion tree has a significant number of nodes at non-trivial depths.
In particular, $\sim$22\% ($\sim$96\%) of records have depth two or greater for IPv4 (IPv6).
Node depth is also significantly long-tailed (with a maximum depth of 9 (15) for IPv4 (IPv6)) indicating that deeper regions of the tree are relatively rare.
Manual inspection of samples of 50 records at depth less than two and two or greater confirmed intuitions about how different depths reflect different types of organizations (\eg deeper networks tend to be smaller regional entities like hospitals or Internet cafes).

\noindent
{\bf Temporal stability of the RIR-level prefix-inclusion tree.}
Next, we investigate how prefix transfers could impact the structure of the RIR-level prefix-inclusion tree using separate public datasets documenting all prefix transfers (\eg~\cite{arinTransfers}).
Due to limited space we provide the details of this investigations in Appendix~\ref{ssec:appx-rir-level}.
Overall, the results indicate that although transfers are a persistent aspect of the address space allocation process, their net effect on address space allocations is relatively small (less that 2\% per year for all RIRs).

\noindent
\textbf{\em Summary.}
Overall, Figures~\ref{fig:whoisNumChildren} and~\ref{fig:whoisDepth} indicate that the explicit prefix-inclusion tree of RIR-level allocations and assignments contributes significant splitting of mass (\ie the tree is wide with many parent prefixes divided in to large numbers of children) and 2 to 4 additional iterations to the conservative cascade process on average.

\subsection{The Organizations: Allocation policy at fine granularity}
\label{ssec:orgs}

Finally, we observe that the particular address regions assigned to ``end-user'' organizations in WHOIS database are {\em not} distributed uniformly at random within their parent prefixes, but exhibit clear clustering in the address space.
This clustering implies that large-scale provider-type organizations (\eg ISPs or multi-national corporations) responsible for parent records have their own internal policies for how to assign address ranges to end-user organizations and that these policies can potentiality be seen as adding additional iterations to the cascade process.
In this section, we seek to quantify the number of additional iterations such clustering may contribute to the cascade process of IP address assignment by {\em (i)} developing a generic ``approximate maximum aggregation'' method to capture this clustering and {\em (ii)} applying this method across the WHOIS dataset.

\begin{figure}[!htb]
    \vspace{-0.2cm}
    \centering
    \includegraphics[width=0.3\columnwidth]{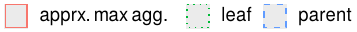}
    \hspace{3em}
    \includegraphics[width=0.37\columnwidth]{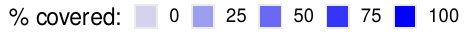}
    
    \begin{subfigure}{0.48\columnwidth}
        \centering
        \includegraphics[width=1.0\columnwidth]{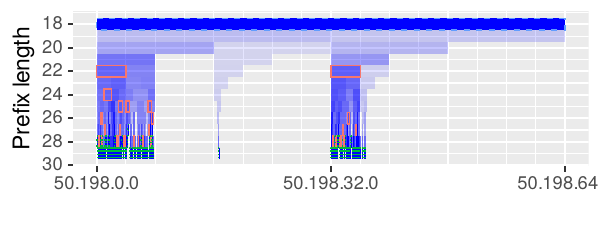}
        \vspace{-0.7cm}
        \caption{Spatial clustering of records (X-axis grid shows /22 blocks).}
        \label{fig:comcastExampleBlocks}
    \end{subfigure}\hfill
    \begin{subfigure}{0.48\columnwidth}
        \centering
        \includegraphics[width=1.0\columnwidth]{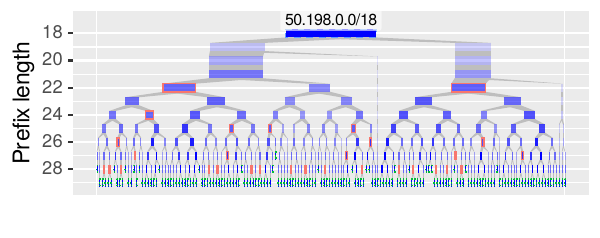}
        \vspace{-0.7cm}
        \caption{Explicit visualization of ``splitting of mass'' in common ancestor tree.}
        \label{fig:comcastExampleTree}
    \end{subfigure}
    \vspace{-0.3cm}
    \caption{Pattern of network allocations in a /18 prefix registered to Comcast.}
    \label{fig:comcastExample}
    \vspace{-0.3cm}
\end{figure}

First, we isolate the network records of these end-user organizations by defining ``leaf'' prefixes as those with zero child records in the prefix-inclusion tree and consider the address-space placement of leaf records within their parent record's prefix.
For example, Figure~\ref{fig:comcastExampleBlocks} visualizes the location of ``leaf'' records that are children of 50.198.0.0/18 (a prefix likely used for Comcast Business customers in Illinois, US) as dotted green rectangles around /28 to /29 on the y-axis.
Manual inspection confirms that these leaves are actual end-user organizations (\eg law firms, schools, or other local businesses).
We observe that the small leaf records form dense clusters around 50.198.0.0 and 50.198.32.0 suggesting at least one additional level of hierarchy in Comcast's organization of this particular /18.
Figure~\ref{fig:comcastExampleTree} explicitly visualizes how the ``mass'' of assignments is split along prefix boundaries by showing all (non-WHOIS) common ancestor prefixes between 50.198.0.0/18 and its (WHOIS) child blocks where the width of each common ancestor prefix is proportional to the number of leaves found in that ancestor.
This clearly exposes the hierarchical structure of the two clusters as two large blocks that begin splitting around /22.

Next, we define the ``\% covered'' metric for each common ancestor prefix as how much of the address space implied by that prefix is covered by a WHOIS record at longer prefix lengths.
For e.g., if there is a WHOIS record at 10.0.0.0/9 and no WHOIS records in 10.128.0.0/9, then 10.0.0.0/8 is 50\% covered.
Figure~\ref{fig:comcastExample} shows ``\% covered'' as the opacity of each prefix's rectangle, indicating that \% covered increases on paths down the common ancestor tree towards the /22 clusters.

Finally, we define ``approximate max aggregate'' (or ``apprx. max agg.'') prefixes as prefixes where the \% covered crosses above a fixed threshold.
Based on the \% covered values in Figure~\ref{fig:comcastExampleBlocks} and similar figures from a sample of other prefixes, we select 51\% covered as a default threshold to capture the visually observed clusters of leaves.
Figure~\ref{fig:comcastExample} outlines the approximate max aggregate prefixes in solid red, showing how this method identifies the /22 clusters obvious in Figure~\ref{fig:comcastExampleBlocks}, but also finer-grained clustering obscured in the figure by the relatively coarse x-axis.

\begin{figure}[!htb]
    \vspace{-0.3cm}
    \begin{minipage}{0.45\columnwidth}
        \centering
        \scalebox{0.9}{
        \begin{tabular}{l | r | r}
        {\bf RIR} & {\bf v4 Max Agg.} & {\bf v6 Max Agg} \\
        \hline
        AFRINIC & 5737 (+4.6\%) & 556 (+1.7\%) \\
        APNIC & 77 501 (+6.2\%) & 2199 (+2.0\%) \\
        ARIN & 326 607 (+10.4\%) & 26 274 (9.2\%) \\
        RIPE & 272 347 (+6.5\%) & 57 569 (+6.5\%)
        \end{tabular}}
        \captionof{table}{Number of approximate max aggregates added (and percentage increase w.r.t. total number of records in each RIR) for each RIR at threshold of 51\% coverage.}
        \label{tab:apprxMaxAggNumbers}
    \end{minipage}\hfill
    \begin{minipage}{0.51\columnwidth}
        \centering
        \includegraphics[width=0.8\linewidth]{fig/global_whois_rir_legend.pdf}
        
        \begin{subfigure}{0.49\linewidth}
            \centering
            \includegraphics[width=1.0\columnwidth]{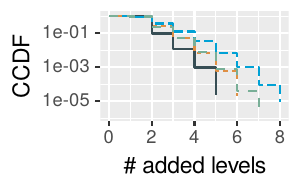}
            \vspace{-0.5cm}
            \caption{IPv4}
            \vspace{-0.2cm}
        \end{subfigure}\hfill
        \begin{subfigure}{0.49\linewidth}
            \centering
            \includegraphics[width=1.0\columnwidth]{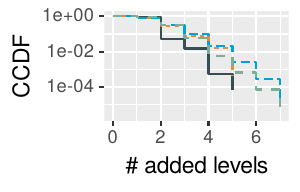}
            \vspace{-0.5cm}
            \caption{IPv6}
            \vspace{-0.2cm}
        \end{subfigure}
        \captionof{figure}{Distribution of \# levels added by approximate max aggregate prefixes over all leaf prefixes of each RIR.}
        \label{fig:whoisOrgAddedLevels}
        \vspace{-0.3cm}
    \end{minipage}
\end{figure}

We apply this method across our dataset of bulk WHOIS records and show how many approximate max aggregate prefixes are found for each RIR in Table~\ref{tab:apprxMaxAggNumbers}.
Table~\ref{tab:apprxMaxAggNumbers} also shows the percentage increase in number of prefixes that would be achieved by adding all approximate max aggregate prefixes to the existing WHOIS records.
Note that due to the inherently large number of ``leaf'' records, these percentage increase numbers may be more substantial than they appear.
In particular, they are actually higher than the percentage of internal (non-leaf) nodes in all cases indicating significant impact on the hierarchical structure.
AFRINIC has a relatively low percentage increase because its WHOIS records are highly organized so that most clusters of leafs already fall smartly into intermediate parent records at several levels.

To quantify the number of cascade iterations potential contributed by approximate max aggregate prefixes, we compute how many approximate max aggregate prefixes are generated on the path between each WHOIS leaf record and its parent record.
Figures~\ref{fig:whoisOrgAddedLevels} shows that this distribution is highly long-tailed for both IP versions with median at 1 and 95-th percentile between 2 and 3 across RIRs.
Excluding AFRINIC (again, due to prevalent covering of internal WHOIS records), in the remaining RIRs between 25\% and 37\% of leaves have two or more approximate max aggregate prefixes between them and their parent WHOIS record (again for both IP versions) indicating the presence of substantial implicit hierarchic organization of leaf records.

\noindent
\textbf{\em Summary.}
Table~\ref{tab:apprxMaxAggNumbers} and Figure~\ref{fig:whoisOrgAddedLevels} indicate that ``leaf'' records associated with particular end-user organizations tend to cluster in well-defined prefix-aligned regions of their parent's address space and that this clustering can be seen as contributing to at least one additional iteration of the conservative cascade IP address allocation process.

\subsection{Establishing the conservative cascade model}

Overall, \S~\ref{ssec:iana}, \S~\ref{ssec:arin-allocs}, and \S~\ref{ssec:orgs} show that at each level of the classic ``three-level story'', internal policies of organizations acting at that level produce highly structured clusters of allocations for the next level.
This demonstrates that a more nuanced, complex, and ultimately ``deeper'' conservative cascade is apparent in the IP address allocation process.
In particular, the IANA-level contributes at least one iteration to the cascade, the RIR-level contributes between 2 and 4 iterations on average, and the organization-level contributes at least one more iteration.
When integrated back into the original ``three-level story'', this yields a cascade with approximately 8 iterations on average.
Moreover, the extremely long-tailed distributions for both the RIR-level and organization-level iteration count estimates suggest that in some cases up to $\sim$10 or up to $\sim$8 additional iterations may be apparent at each of these levels respectively.
These observations lead to the conclusion that the ``physical explanation'' for the structure of observed IP addresses lies in the non-trivial conservative cascade responsible for splitting the ``mass'' of address assignments hierarchically across the prefix-level structure of the address space.

\section{Conservative Cascade as a Generative Model for IP Addresses}
\label{sec:generative}

Having established that a mechanism consistent with a conservative cascade process is responsible for allocation of IP address ranges to different end-users in the previous section, we now turn to the question of how to leverage such a cascade process to directly generate sets of IP addresses with predicable structural properties.
First, to formalize the conservative cascade model, for each $l \geq 0$, we identify the $l$-th dyadic partition of $[0,1) \subset \mathcal{R}$ with the set of binary sequences $x_{1}, x_{2}, \dots, x_{l}$ of length $l$ where $x_{i} \in \{0, 1\}$ for $1 \leq i \leq l$.
If $x = (x_{1}, x_{2}, \dots, x_{l})$ denotes such a binary sequence of length $l$, then the corresponding dyadic interval given by $x$ is $[\sum_{1 \leq i \leq l} x_{i}2^{-i}, \sum_{1 \leq i \leq l} x_{i}2^{-i} + 2^{-l})$.
In terms of IP addresses, each binary sequence $x = (x_{1}, x_{2}, \dots, x_{l})$ of length $l$ maps to a unique prefix of length $l$, where the prefix is comprised of all IP addresses whose $i$-th bit is given by $x_{i}$ ($1 \leq i \leq l$).
In theory, as $l \to \infty$, $x$ converges to a real number in $[0,1) \subset \mathcal{R}$.
However, in practice, because of the discrete and finite nature of the IP address space, this mathematical limit does not exist, which in turn  necessitates studying the ``large-$l$ asymptotics"  over a range of $l$-values that has a hard upper limit of $32$ for IPv4 or $128$ for IPv6.

Next, let $W$ be a fixed random variable that takes values in $[0,1)$, has mean 1/2, and is symmetric about its mean.
The conservative cascade model with generator $W$ can be viewed as a set of random variables $\{W_{l,j}\ :\ l>0 $ and $1 \leq j \leq 2^l\}$  where {\em (i)} the $W_{l,j}$'s corresponding to the $l$-th dyadic partition of $[0,1)$ are independent from those corresponding to the $k$-th dyadic partitions (for all $k \neq l$) and have the same distribution as $W$, and {\em (ii)} the $W_{l,j}$'s within the $l$-th dyadic partition of $[0,1)$ ($l>0$) have a dependence structure given by $W_{l, j + 1} = 1 - W_{l, j}$ for odd $j$'s (note that because of the properties of $W$, $W_{l,j+1}=1-W_{l,j}$ are identically distributed as $W$). For each  $l>0 $, the $W_{l,j}$'s determine how much ``mass" is distributed from each parent in the $(l-1)$-th dyadic partition of $[0,1)$ to its two children in the $l$-th dyadic partition and ensure that this distribution is done in a mass-preserving manner. By the independence of the $W_{l,j}$'s between different dyadic partitions or levels $l$, the measure of ``mass'' for a given $x$ at level $l$ is
\[
\mu_{l}(x) = Z \cdot \prod_{1 \leq i \leq l}W_{i,j_{i}}
\]
where $Z$ represents the total mass of the cascade and $j_{i} = 1 + \sum_{1 \leq k \leq i} x_{k}2^{i - k}$ specifies the prefixes at levels $i \leq l$ that contain $x$ (\ie the branches in the binary tree).
In theory, the measure $\mu_{\infty}$ obtained by taking the mathematical limit of $\mu_{l}(x)$ as $l \to \infty$ for all $x \in [0,1) \subset \mathcal{R}$ constitutes a ``realization'' of this conservative cascade model and can be described in terms of well-defined theoretical properties such as the fractal dimension of its support~\cite{ossiander2000statistical,gilbert-etal:2003}.
However, in practice, the inherently finite and discrete nature of the IP address space again prevents us from leveraging genuinely asymptotic limits and requires us to focus instead on ``large-$l$ asymptotics" that have a hard upper limit of  $l=32$ for IPv4 or $l=128$ for IPv6.
This produces a ``discrete realization" of the conservative cascade that details how the (discrete) total of $Z$ observed IP addresses are distributed across the IP space at level $l=32$ (\ie which /32's have mass 1).

This construction suggests direct approaches to both {\em (i)} estimating the distribution of $W$ from an observed set of IP addresses and {\em (ii)} synthetically generating new sets of IP addresses by repeatedly drawing samples from a particular distribution of $W$---ideas also suggested in prior work~\cite{sommers2012efficient,barford2006toward}.
For the former, given a discrete realization of the conservative cascade, %
we simply consider the empirical distribution of the values taken by each $W_{l,j}$ and given by $w_{l,j} = \mu_{l}(x_{1},\dots,x_{j}) / \mu_{l-1}(x_{1},\dots,x_{j-1})$.
For the later, we start with a given number of addresses (\ie our address ``mass'' $Z$), generate random variates $W_{l,j}$ (for odd $j$), and successively divide our addresses at each level $l$ until we reach the hard upper limit of 32 (IPv4) or 128 (IPv6).
Unfortunately, due to the finite and discrete nature of the IP space, both of these approaches run into difficulties ({\em not} identified in prior work) and require modifications that are detailed below.
These modifications are intended to not only result in practical solutions but also preserve the theoretical underpinning for conservative cascade models.

\subsection{Fitting to observed addresses}
\label{ssec:fitting}

\begin{figure}[!htb]
    \vspace{-0.6cm}
    \centering
    \begin{minipage}{0.48\columnwidth}
        \includegraphics[width=1.0\linewidth]{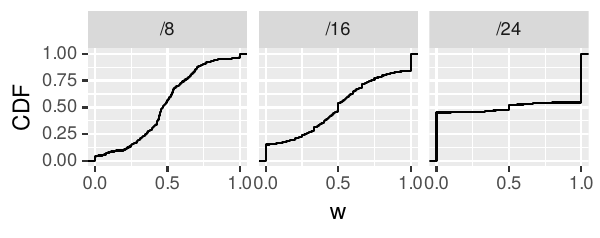}
        \vspace{-0.7cm}
        \caption{Distribution of the raw values of $w_{l,j}$ for the CAIDA500k dataset at various prefix lengths.}
        \label{fig:weigthsCaidaRawSmall}
    \end{minipage}\hfill
    \begin{minipage}{0.48\columnwidth}
        \centering
        \includegraphics[width=1.0\linewidth]{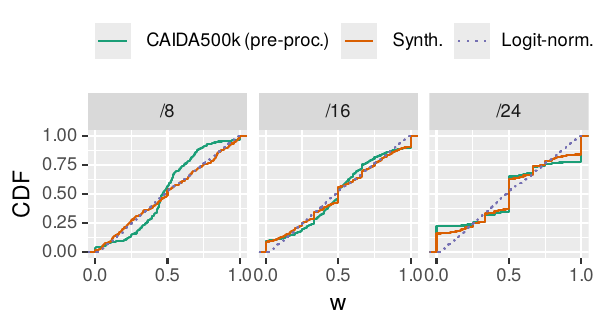}
        \vspace{-0.7cm}
        \caption{Comparison of $w_{l,j}$ from the pre-processed CAIDA500k dataset, a synthetic cascade realization (as described in \S~\ref{ssec:generating}), and the theoretic best-fit Logit-normal distribution ($\sigma = 1.61$).}
        \label{fig:weigthsFitSmall}
    \end{minipage}
    \vspace{-0.5cm}
\end{figure}

To illustrate the process of directly fitting a distribution for $W$ based on observed values $w_{l,j}$, we compute the values $w_{l,j}$ for the real-world CAIDA500k dataset and show their distribution for three example prefix lengths (in particular $l = 8, 16, 24$) in Figure~\ref{fig:weigthsCaidaRawSmall}.
These distributions are almost exactly symmetric (\ie mean 0.5) and closely resemble well-known distributions that take values in $[0,1)$ such as the Logit-normal or Beta distribution, especially at short to medium prefix lengths (\eg /8 - /16).
Based on this finding and for the purpose of concreteness, we consider in this paper symmetric Logit-normal distributions due to their single, easily-estimated parameter ($\sigma$), though our methods could easily be applied to other $[0,1)$ distributions such as symmetric Beta distributions.

Despite suggesting a general distributional shape, Figure~\ref{fig:weigthsCaidaRawSmall} also exposes several salient challenges in fitting the $w_{l,j}$ to continuous parametric models like the Logit-normal distribution.
Due to space limitations we summarize the pre-processing steps we developed to address these challenges here and provided detailed analysis in Appendix~\ref{ssec:appx-fittingDetails}.

\squishlist
    \item 
    Sparseness at longer prefix lengths leads to a large number of prefixes with only a single address, forcing $w_{l,j}$ to be 0 or 1. Since such prefixes cannot provide any information about how mass would be split, we ignore them when performing the Logit-normal fit.
    \item
    Cases where a prefix has more than one address but the entire address ``mass'' moves exclusively to the left or right child also result in $w_{l, j}$ = 0 or 1. When $w_{l,j}$ is 0 we replace it with 1 / $2n$ and when $w_{l,j}$ is 1 we replace it with 1 - 1 / $2n$ where $n$ is the number of addresses in the ``parent'' prefix. This corrects for ``rounding'' effects in translation from an imagined continuously-distributed $W$ to the distinct $w_{l,j}$ we observe.
    \item
    We only use values of $w_{l,j}$ from $8 \leq l \leq 16$ (\ie only prefixes between /8 and /16) to avoid the noisy behavior at shorter prefix lengths and the more pronounced perturbations by discrete values at longer prefix lengths.
\squishend

By applying these pre-processing steps and the usual standard-deviation estimator to the Normal-transformed $w_{l,j}$'s, we fit the CAIDA500k data's generator $W$ to a Logit-normal distribution with parameter $\sigma = 1.61$.
Figure~\ref{fig:weigthsFitSmall} shows the distribution of the pre-processed $w_{l,j}$ for the CAIDA500k data (solid green), a synthetic finite discrete conservative cascade as described in the next section (solid orange), and a numerically computed Logit-normal distribution with $\sigma = 1.61$ (dotted purple).
Though it is challenging to directly assess goodness of fit due to
discrepancies between the theoretic Logic-normal distribution and the empirical data (with perturbations from its inherent discreteness), we note that it is qualitatively close, especially for longer prefix lengths (\eg around /16).
Moreover, the distributions of data synthesized with the adapted version of the conservative cascade (discussed next) come even closer to the empirical CAIDA500k data, especially at longer prefix lengths (where the sample size is inherently larger).

\subsection{Generating finite discrete sets of IP addresses}
\label{ssec:generating}

Given the methods for estimating a parametric distribution for $W$ discussed above, we now turn to the matter of using an estimated distribution as a generator in a conservative cascade to produce a synthetic set of IP addresses.
The inputs to this method are {\em (i)} a distribution $W$ on $[0,1] \subset \mathcal{R}$ (cascade generator) from which we can draw random variates $w$ and {\em (ii)} a number $n = n_{0,0}$ of how many IP addresses to generate (\ie the total ``mass'' of the synthetic cascade).
Starting from the root of the prefix tree with $l = 0$ (\eg 0.0.0.0/0 for IPv4), this method works by generating a random variate $w_{l,j}$, dividing the mass of the parent prefix $n_{l - 1, j}$ among its two child prefixes by assigning them the mass $n_{l, 2j} = [w_{l,j} \cdot n_{l-1, j}]$ and $n_{l, 2j+1} = [(1 - w_{l,j}) \cdot n_{l-1,j}]$, respectively,\footnote{Here, $[ ]$ denotes rounding to the nearest integer.} and recursively repeating this process until all mass has been distributed to $l = 32$ (for IPv4) or $l = 128$ (for IPv6).

Despite the method's apparent simplicity, there are two problematic cases---where the number of addresses assigned by $w_{l,j}$ to a child prefix is larger than the total number of addresses that this child prefix can possibly contain---that need attention.
The first case is due to the fact that a prefix with $k$ ``host'' bits can contain at most $2^k$ distinct addresses.
For example, a /24 IPv4 prefix can contain at most 256 distinct addresses, or in the extreme case, a /32 IPv4 prefix can only contain one distinct address.
The second case concerns the reserved address ranges that IANA designates for special usage such as internal networking or multicast protocols.
To generate realistic synthetic sets of ``observed'' addresses, these regions should contain exactly zero addresses since they are (typically) not routed on the public Internet and hence never observed.

To deal with these cases, we associate a capacity with each prefix based on prefix length and well-known reserved prefixes.
When dividing a parent prefix's mass among its two child prefixes in the synthetic cascade, we ``spill over" any addresses above the capacity of one of child's prefix to the other child's prefix.
For example, suppose we need to assign $n_{l-1, j} = 100$ addresses to children $n_{l, 2j}$ and $n_{l, 2j+1}$ with $w_{l, j} = 0.8$ and child capacities of 64 and 256 respectively.
Because $[w_{l,j}\cdot n_{l-1,j}] = [100 \cdot 0.8] = 80$ is larger than the capacity of the first child, we set $n_{l, 2j} = 64$ and $n_{l, 2j + 1} = 16 + 20 = 36$, spilling the extra addresses over to the child with extra capacity.
Note that as long as the total capacity of the cascade (\ie the capacity of 0.0.0.0/0 or ::/0) is greater than or equal to the number of addresses to generate, this method will always generate exactly the required number of addresses while respecting the given capacity constraints.

In Figure~\ref{fig:weigthsFitSmall}, the solid orange line shows the distribution of the weights $w'_{l,j}$ after applying the methods described here for a synthetic finite discrete conservative cascade with a total of $n_{0,0} = 500k$ distinct IPv4 addresses and the fitted Logit-normal generator $W$ described previously.
The distribution of the synthetic $w'_{l,j}$'s closely matches the theoretical Logit-normal distribution (dotted purple) at short prefix lengths (\eg /8 or /12) and inherits the same increased variance compared to CAIDA500k (solid green).
However, at longer prefix lengths (\eg /20 or /24) the synthetic $w'_{l,j}$ begin featuring nearly identical sharp modes (\eg at 0, 0.5, and 1) as the CAIDA500k $w_{l,j}$.
This confirms both that these perturbations are likely the result of the finite and discrete nature of IP addresses and that our approaches for dealing with this IP space-specific constraint in the construction of synthetic conservative cascades yields results that are in close agreement with those obtained from real-world data (\ie observed IP addresses in measured Internet traffic traces).
In Figure~\ref{fig:generaticPicture} of Appendix~\ref{ssec:appx-pictorialSynthEvidence} we also confirm that the IP addresses generated with this method yield similar structural properties as shown for the CAIDA500k dataset in Figure~\ref{1Dfigure:compare:caida}.

\noindent
\textbf{\em Summary.}
In this section we showed how the (infinite continuous) conservative cascade model can be adapted to (finite discrete) sets of IP addresses and demonstrated that the cascades constructed with this adaptation qualitatively match real-world IP address structure.
We will continue to refer to synthetic conservative cascades generated using this method with different logit-normal variance parameters $\sigma$ to illustrate how the variance of the generator and the finite discreteness of IP addresses impact the clustering and scaling properties of the resulting address sets.

\section{Multifractal IP Address Structure: Fait Accompli }
\label{sec:evidence}

Given the correspondence between IP address allocation policies and the conservative cascade model (\S~\ref{sec:physical-explanation}), the ability to infer the distribution of the cascade generator from a single discrete realization of the model (\ie real-world data given in the form of a set of observed IP addresses, \S~\ref{ssec:fitting}), and the means to use the inferred generator to synthetically generate new discrete cascade realizations (\S~\ref{ssec:generating}), we next turn to multifractal analysis to characterize the statistical properties of the sets of observed IP addresses produced by these cascade models.

\subsection{Using the method of moments ``out-of-box"}

Returning to the general-case notation of \S~\ref{sec:background} and Appendix~\ref{ssec:multifracatal-analysis}, applying the method of moments (as described in~\cite{evertsz1992multifractal,salat2017multifractal}) involves {\em (i)} computing partition functions $Z(r, q) = \sum_{x} \mu(B(r, x))^{q}$ for a range of ball sizes $r$ and ``moments'' $q\in\mathcal{R}$, {\em (ii)} examining the partition functions' scaling behavior as $r \to 0$ by determining for each $q$ whether or not a plot of $\log(Z(r, q))$ against $\log(r)$ forms a straight line, and {\em (iii)} estimating the structure function $\tau(q)$ as the slope of this $q$-dependent straight line.
Connecting to the specific case of IP addresses and the notation of \S~\ref{sec:generative}, we compute $Z(l,q) = \sum_{x\ :\ |x| = l} \mu_{l}(x)^{q}$ (where the sum is taken over all prefixes of length $l$) and consider the plot of $\log_{2}(Z(l,q))$ against $-l$ (because $r = 2^{-l}$).
Finally, because of effects of finite discreteness discussed below,
we select an intermediate range of prefix lengths where the linear behavior of $\log_{2}(Z(l,q))$ is most pronounced to estimate $\tau(q)$ and ignore shorter and longer prefix lengths.
We also consider a limited range of $q$-values as suggested by several efforts~\cite{lashermes2004new,ossiander2000statistical}.

\begin{figure}[!htb]
    \vspace{-0.3cm}
    \centering
    \includegraphics[width=0.65\linewidth]{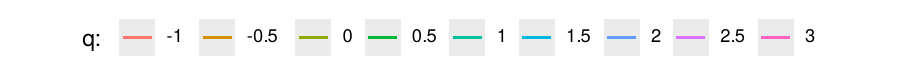}
    
    \begin{subfigure}{0.25\textwidth}
        \includegraphics[width=0.9\columnwidth]{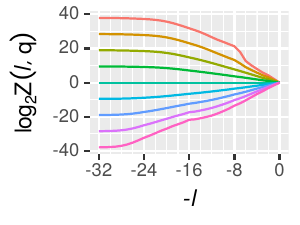}
        \vspace{-0.3cm}
        \caption{CAIDA500k}
        \label{fig:partitionFunctions:caida}
    \end{subfigure}~
    \begin{subfigure}{0.25\textwidth}
        \includegraphics[width=0.9\columnwidth]{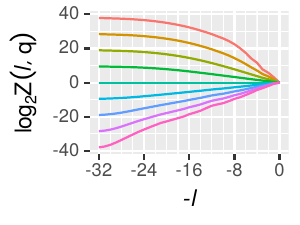}
        \vspace{-0.3cm}
        \caption{Synth.}
        \label{fig:partitionFunctions:synth}
    \end{subfigure}~
    \begin{subfigure}{0.25\textwidth}
        \includegraphics[width=0.9\columnwidth]{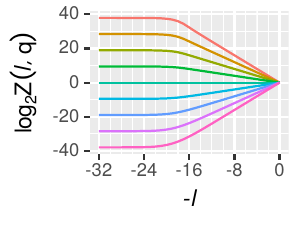}
        \vspace{-0.3cm}
        \caption{Uniform500k}
        \label{fig:partitionFunctions:uniform}
    \end{subfigure}
    \vspace{-0.3cm}
    \caption{``Partition function'' plots of $\log_{2}(Z(q,l))$ vs. $-l$ for real-world (CAIDA500k), synthetic (Synth.), and uniform random (Uniform500k) data.}
    \label{fig:partitionFunctions}
    \vspace{-0.6cm}
\end{figure}

Figure~\ref{fig:partitionFunctions} shows these plots for CAIDA500k, the best-fit synthetic conservative cascade from \S~\ref{sec:generative} with $\sigma = 1.61$ (Synth.), and Uniform500k.
As expected, the partition function for Uniform500k shows linear behavior (from /0 through around /20 when sparsity effects take over) which is indicative of mono-fractal scaling.
For CAIDA500k and the synthetic data, several different behaviors are observed over different prefix length ranges.
Observing this more complex behavior in these cases first calls for an investigation of its root causes (\S~\ref{ssec:effectsOfDiscrete}), and then using this understanding to develop an improved version of the existing method of moments that is amenable to analyzing discrete and finite settings such as IP spaces (\S~\ref{ssec:improvedMoM}).

\subsection{IP spaces are finite and discrete: Ramifications for inferring scaling behaviors}
\label{ssec:effectsOfDiscrete}

To understand why the partition functions of Figure~\ref{fig:partitionFunctions} exhibit distinct nonlinearity for short (\eg /8) and long (\eg /16 through /24) prefix lengths, we first hypothesize that these nonlinear behaviors are artifacts of ``edge-cases" arising from the finite and discrete nature of the IP space.
This allows us to view IP addresses as discrete approximations of suitably selected continuous counterparts in order to use the same tools and methods that have been developed for the continuous case, albeit modified for or tailored to the discrete case.
To explore this hypothesis, we use our finite discrete generation method to synthesize additional conservative cascades with generators that have a range of different variance parameters.

Due to limited space, we discuss the details of this analysis in Appendix~\ref{appx:analysisDetails} and only summarize the final takeaways here.
\squishlist
    \item 
The nonlinearity at short prefix lengths impacts negative values of $q$ associated with smaller prefixes and is correlated with the emergence of prefixes with only a single IP address (which essentially cease to scale in any meaningful way).
    \item 
The nonlinearity at long prefix lengths impacts positive values of $q$ associated with larger prefixes and is correlated with ``spillover'' induced by the inherently limited capacity at long prefix lengths (\eg a /24 prefix can only contain 256 distinct addresses).
\squishend
These two observations indicate exactly the ``edge-cases'' where our view of IP addresses as discrete approximations breaks down requiring modifications to the standard infinite continuous multifractal analysis tools.

\subsection{An improved method-of-moments for analyzing IP address structure}
\label{ssec:improvedMoM}

The explanations for the observed nonlinearity of the partition functions described in the previous subsection point to two types of ``anomalous'' behavior that must be taken into consideration when applying the existing method-of-moments analysis to sets of IP addresses.
At a fundamental level, these explanations suggest that to avoid the considered edge cases caused by either the short- or long-length prefixes, we must carefully select a range of prefix lengths between the left and right tails of the available spectrum of prefix lengths to capture to the best of our ability the ``true" scaling behavior of a given IP address set.
To this end, we focus on prefix lengths between /8 and /16 since this range appears to be safely between the edge-cases based on our analysis in Appendix~\ref{appx:analysisDetails}. %

Next, we note that the first explanation (regarding the short prefix lengths) also raises a more subtle concern stemming from the nontrivial existence of prefixes with a single address.
From the multifractal scaling perspective, these single-address prefixes imply that if for some point $x$, $\mu(B(r, x)) = 1$, then $\mu(B(r',x)) =1$ for all $r`<r$, implying that the local Holder exponent $\alpha(x) = 0$ which violates the general property $0 < \alpha_{min}$ (see, \eg~\cite{evertsz1992multifractal}).\footnote{Some theoretical works~\cite{mandelbrotAnomalous,hananLeftsided} describe an ``anomalous'' or ``one-sided'' type of multifractal scaling where $\alpha_{min}$ can be zero, but there is presently no clear physical justification for application of these theories to IP addresses and we leave deeper exploration to future work.}
To estimate the actual scaling structure, we should therefore ignore these single-address prefixes and focus only on cases where splitting of address ``mass'' contributes to nontrivial scaling.
However, simply removing all single-address prefixes is not feasible in the traditional method-of-moments approach because substantial address ``mass'' could be lost at each increasing prefix length, making it impossible to identify a sufficiently wide range of prefix lengths for obtaining meaningful and consistent estimates of the slopes of the partition functions (\eg with the standard least-squares estimator).

To address this concern, we leverage an alternative estimator of the slope of the partition functions that was introduced in~\cite{ossiander2000statistical} and is tailored to a general class of cascade-based multifractals.
In particular, we use the estimator
\[
\tilde{\tau}_{l}(q) = -\log_{2}\left(Z(q, l + 1) / Z(q, l)\right) = \log_{2}(Z(q, l)) - \log_{2}(Z(q, l + 1)).
\]
Instead of relying on fitting to a range of prefix lengths, $\tilde{\tau}_{l}(q)$ simply takes the difference of $\log_{2}(Z(q,l))$ between two consecutive prefix lengths.
Remarkably, \cite{ossiander2000statistical} shows that for a general type of multiplicative cascades of which the conservative cascade models considered here is a special case, this estimator is not only asymptotically consistent for a range of $q$ values as $l \to \infty$ but also satisfies a central limit theorem (\ie asymptotic distribution is Gaussian) and allows the computation of confidence intervals based only on observed statistics.

Beyond these useful properties, the reliance on only two adjacent prefix lengths at a time allows us to apply $\tilde{\tau}(q)$ ``around'' single-address prefixes by filtering all single address prefixes at length $l$, but including any single-address child prefixes generated at length $l+1$.
Since the physical IP space constraints prevent us from taking the limit of $\tilde{\tau}(q)$ as $l \to\infty$, we instead select a range of prefix lengths (as discussed above), apply $\tilde{\tau}_{l}(q)$ per prefix length, and average the results to obtain a ``pseudo-asymptotic" estimate.
In particular, given a sequence of prefix lengths $L = l_{1}, l_{2}, \dots, l_{n}$, assuming independence of the fluctuations around $\tau(q)$ for different q-values, and leveraging the central-limit theorem from~\cite{ossiander2000statistical}, we consider the averaged estimator
$\tilde{\tau}(q) = (1/n)\sum_{i = 1}^{n} \tilde{\tau}_{i}(q)$
which is normally distributed around $\tau(q)$ with variance $(1/n^{2})\sum_{i = 1}^{n}D^{2}_{i}(q)$ where the $D^{2}_{i}(q)$'s are computable quantities that are explicitly defined in~\cite{ossiander2000statistical}.

\subsection{Multifractal IP address structure as a new invariant of Internet traffic}
\label{ssec:multifractalAsNewInvar}

\begin{figure}[!htb]
    \vspace{-0.3cm}
    \centering
    \begin{subfigure}{0.3\textwidth}
        \includegraphics[width=0.9\columnwidth]{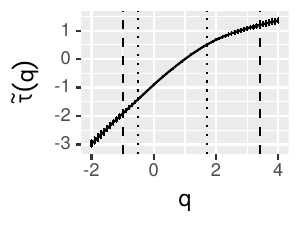}
        \vspace{-0.3cm}
        \caption{CAIDA500k}
        \label{fig:partitionFunctions:caida}
    \end{subfigure}~
    \begin{subfigure}{0.3\textwidth}
        \includegraphics[width=0.9\columnwidth]{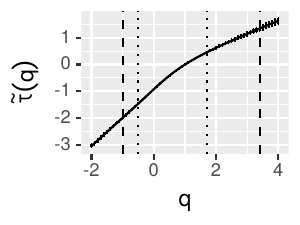}
        \vspace{-0.3cm}
        \caption{Synth.}
        \label{fig:partitionFunctions:synth}
    \end{subfigure}~
    \begin{subfigure}{0.3\textwidth}
        \includegraphics[width=0.9\columnwidth]{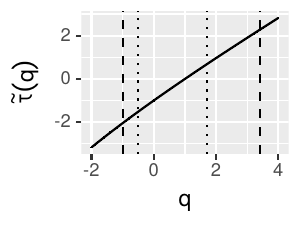}
        \vspace{-0.3cm}
        \caption{Uniform500k}
        \label{fig:partitionFunctions:uniform}
    \end{subfigure}
    \vspace{-0.3cm}
    \caption{$\tilde{\tau}(q)$ (the estimated ``structure functions'') for real-world (CAIDA500k), synthetic (Synth.), and uniform random (Uniform500k) data with 95\% confidence intervals. Dashed lines show the critical range for convergence, dotted lines show the critical range for normalicy.}
    \label{fig:structureFunctions}
    \vspace{-0.3cm}
\end{figure}

Using $\tilde{\tau}(q)$, we estimate the ``structure functions'' of all sets of addresses introduced in \S~\ref{sec:background} and use as examples in Figure~\ref{fig:structureFunctions} the same three sets considered in Figure~\ref{fig:partitionFunctions}.
Leveraging another set of theoretical results from~\cite{ossiander2000statistical} and using them for a conservative cascade with a Logit-normal generator with $\sigma = 1.61$, we also estimate and plot the critical range of $q$ values for which the asymptotic consistency (-1.0 to 3.4, dashed lines) and asymptotic Gaussianity (-0.5 to 1.7, dotted lines) properties of $\tilde{\tau}(q)$ hold.
We note that in agreement with the ``visual proof'' in Figure~\ref{1Dfigure:compare}, the observed nonlinearity of the CAIDA500k and synthetic dataset structure functions implies consistency with multifractal scaling while the shown linearity of the Uniform500k structure functions implies consistency with mono-fractal scaling. Morevover, the nonlinear shapes of the structure functions for CAIDA500k and the synthetic dataset are consistent with theory in that they are increasing, concave-down functions, and their knees (which indicate the presence of a range of positive scaling exponents $\alpha$) are well within the critical regions.

We also show in Table~\ref{table:gen-dim} estimates of the generalized dimensions $D_0$, $D_1$, and $D_2$ for the same three datasets and include as a sanity check whether or not the estimated $\tau(1)$-value is zero in agreement with theory. The information in this table succinctly quantifies the visible differences between the left and right plots in Figure~\ref{1Dfigure:compare}. On the one hand, the multifractal data on the left (CAIDA550k) does not fill the physical space (\ie $D_0 < 1$), is highly unevenly distributed (\ie $D_1$ is small), and exhibits significant scattering (\ie $D_2$ is small).\footnote{We also checked (not shown) that the obtained results are largely insensitive w.r.t. data size.}
On the other hand, the generalized dimensions for the monofractal data on the right (Uniform500k) are degenerate in the sense that $D_0 = 1$ (\ie the data fill the unit interval); $D_1$ is large (\ie very evenly distributed); and $D_2$ is large (\ie more compactness).

\begin{table}[htb!]
\vspace{-0.2cm}
\centering
\scalebox{0.8}{
\begin{tabular}{ | c | c | c | c | c | }
\hline
ToI Instances & \makecell{$D_0$ (fractal dim.)} & \makecell{$D_1$ (information dim.)} & \makecell{$D_2$ (correlation dim.)} & \makecell{$\tilde{\tau}(1) = 0?$} \\
 \hline
 \hline 
 \textit{CAIDA500k} & 0.90$\pm$0.001  & 0.81 & 0.68$\pm$0.009 & \checkmark  \\  
 \hline 
 \textit{Synth.} & 0.92$\pm$0.001  & 0.74 & 0.63$\pm$0.012 & \checkmark  \\  
 \hline
 \textit{Uniform500k} & 1.00$\pm$0.000 & 0.99 & 0.96$\pm$0.000  & \checkmark  \\  
 \hline
 \hline
 \textit{CAIDA\_dirA\_v6} & 0.29$\pm$0.003 & 0.16 & 0.12$\pm$0.014 & \checkmark \\
 \hline
\end{tabular}}
\vspace{.1in}
\caption{Estimated generalized dimensions and standard deviation based on $\tilde{\tau}(q)$ where applicable. Results for the rest of our repository of ToI are in Table~\ref{table:analysis-results}.}
\label{table:gen-dim}
\vspace{-0.8cm}
\end{table}

\begin{figure}[!htb]
    \vspace{-0.3cm}
    \centering
    \begin{subfigure}{0.3\textwidth}
        \includegraphics[width=0.9\columnwidth]{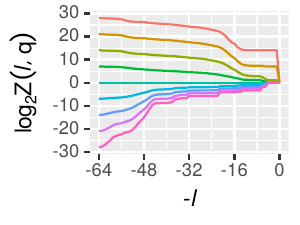}
        \vspace{-0.3cm}
        \caption{Partition functions}
        \label{fig:initialv6Analysis:partition}
    \end{subfigure}~
    \begin{subfigure}{0.3\textwidth}
        \includegraphics[width=0.9\columnwidth]{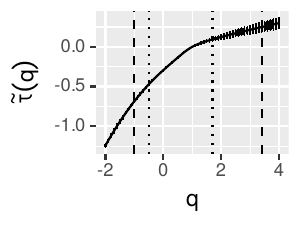}
        \vspace{-0.3cm}
        \caption{Structure function}
        \label{fig:initialv6Analysis:struct}
    \end{subfigure}~
    \vspace{-0.3cm}
    \caption{Partition and structure functions for $\sim$17k IPv6 addresses in CAIDA\_dirA\_v6.}
    \label{fig:initialv6Analysis}
    \vspace{-0.2cm}
\end{figure}

Next, we illustrate the similarities/differences of multifractal scaling in observed IPv6 addresses by considering the CAIDA\_dirA\_v6 dataset made from the $\sim$17k distinct IPv6 addresses in one hour of direction A of the CAIDA trace.
Note that for privacy the lower 64 bits of IPv6 addresses are set to zero in this data set so our analysis is limited to the upper 64 bits.
Figure~\ref{fig:initialv6Analysis} shows the partition and structure functions for CAIDA\_dirA\_v6 considering the medium-range scaling behavior between /20 and /44.
We also provide the generalized dimensions for CAIDA\_dirA\_v6 in Table~\ref{table:gen-dim}.

The strongly non-linear structure function in Figure~\ref{fig:initialv6Analysis:struct} indicates that CAIDA\_dirA\_v6 is also consistent with multifractal scaling.
Given the relative scarcity of observed IPv6 addresses (compared to $\sim$2.4M IPv4 addresses) in the CAIDA trace, it is hard to make any definitive observations about differences in scaling behavior.
In particular, the smaller fractal dimension of CAIDA\_dirA\_v6 (0.29) is a direct reflection of the sparser population of IPv6 space (\eg as shown in Figure~\ref{fig:iana_v6Timeline}) and the smaller information and correlation dimensions (0.16 and 0.12 respectively) reflect increased unevenness and scatteredness compared to the IPv4 case, possibly resulting from the longer-tailed distribution of allocation node degree (\eg as shown in Figure~\ref{fig:whoisNumChildren:v6}).

The results for all datasets mentioned in Section 2 and listed in Table~\ref{table:list_of_datasets} in Appendix~\ref{appx:traceRepo} are summarized in Table~\ref{table:analysis-results} in Appendix~\ref{appx:resultsSummary}. The results of this comprehensive sample support our main finding that the obtained evidence in favor of a multifractal structure of observed IP addresses in measured network traffic is not limited to a specific time, place, or type of traffic. All measured traces produce partition and structure function plots (not shown) that are similar to Figures~\ref{fig:partitionFunctions} and~\ref{fig:structureFunctions}, respectively, and yield generalized dimension estimates that are similar to those obtained for CAIDA500k and distinctly not like those obtained for Uniform500k (see Table~\ref{table:analysis-results} in Appendix~\ref{appx:resultsSummary}). In view of our empirically validated physical explanation provided in Section 3, these results are fully expected and confirm multifractal IP address structure to be an invariant of measured traffic, a facet of behavior of measured Internet traffic which has been empirically shown to hold in a wide range of environments and include such properties as diurnal patterns of activity, (asymptotic) self-similar scaling of temporal traffic rate processes, and heavy-tailed distributions for protocol-related entities such as TCP connection or IP flow duration and size (\eg see~\cite{paxson-floyd:ton2001} and references therein).

\noindent
\textbf{\em Summary.}
In this section we adapted a robust estimator of the structure function $\tilde{\tau}(q)$ to handle edge-cases arising from the finite discrete IP address structure and used this estimator to demonstrate strong statistical evidence for multifractal scaling across a wide range of sets of observed IP addresses drawn from real-world Internet traffic.

\section{Application: Anomaly Address Detection}
\label{sec:implications}

In this section we consider the task of detecting anomalous IP addresses in observed Internet traffic as an illustrative concrete example of how the quantitative analysis of address structure discussed in the previous sections can be put to practical use.
Organizations such as ISPs and other service providers regularly assess the reputation of addresses entering their network and make proactive decisions based on this assessment, such as blocking addresses associated with spam, phishing, low-rate protocol-conforming DDoS, or other malicious activities~\cite{li2021clairvoyance,kogan2021private}.
Such decisions serve as a critical first line of defense against known and suspected adversaries and are efficient to implement using common access-control mechanisms.

Unfortunately, present-day approaches which primarily use blocklists~\cite{blocklistde,firehol,ipsum,proofpoint} to establish reputation suffer from two fundamental limitations and cannot provide a useful assessment of an arbitrary address's degree of anomalousness or ``anomalous score''.
First, blocklists typically only provide a ``negative'' reputation for a small number of known malicious addresses or prefixes.
To quantify this limitation, Table~\ref{tab:blocklistCoverage} shows the number of addresses included in several well-known public blocklists and the fraction of the total address space they cover.
Overall, these lists cover very small slices of the total space of possible addresses with maximum coverage (provided by FireHOL) only accounting for $\sim$14\% of possible addresses.
Second, blocklists are often sourced from third-party ``threat-intelligence'' services, forming a centralized point of uncertainty and failure~\cite{feal2021blocklist}, and their static nature prevents them from dealing with dynamic address behaviors caused by operations such as NATing or address reuse~\cite{ramanathan2020quantifying}.

\begin{wraptable}{R}{0.45\columnwidth}
    \centering
    \scalebox{0.8}{
    \begin{tabular}{l|r|r}
    {\bf Block-list name} & {\bf \# addresses} & {\bf \% coverage} \\
    \hline
    Blocklist.de~\cite{blocklistde} & $\sim$24k & $\sim5.7 \cdot 10^{-4}$ \\
    FireHOL~\cite{firehol} & $\sim$612M & $\sim1.4 \cdot 10^{+1}$ \\
    IPsum~\cite{ipsum} & $\sim$205k & $\sim4.8 \cdot 10^{-3}$ \\
    Proofpoint~\cite{proofpoint} & $\sim$16M & $\sim3.6 \cdot 10^{-1}$
    \end{tabular}}
    \caption{Summary of publicly-available IPv4 address ``block-lists'', the total number of distinct addresses covered by each, and the equivalent percentage covered compared to the entire valid IPv4 address range.}
    \label{tab:blocklistCoverage}
    \vspace{-0.5cm}
\end{wraptable}

To address these limitations of the list-based approach, we develop a more general notion of ``anomalous score'' associated with incoming IP addresses and based on detecting changes in their structure, rather than a centralized threat-intelligence service.
In particular, given a stream of the distinct IP addresses observed entering a particular network, we assign to {\em each} address an anomalous score that quantifies the level of suspicion that should be associated with that address.
This score is meant to complement existing blocklist-based approaches, providing network operators with additional structural context information about the large number of incoming addresses not included in existing blocklists.

\begin{figure}[!htb]
    \centering
    \vspace{-0.3cm}
    \includegraphics[width=0.95\linewidth]{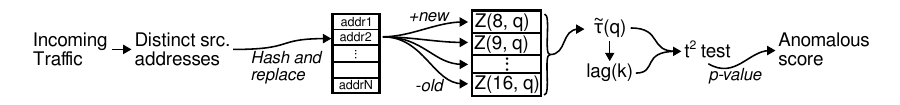}
    \vspace{-0.3cm}
    \caption{Approach to detecting anomaly addresses by applying a lagged $t^2$-test over the sequence of incoming distinct addresses.}
    \label{fig:anomDetectionOverview}
    \vspace{-0.4cm}
\end{figure}

To compute this anomalous score, we adapt the structure function estimator $\tilde{\tau}(q)$ to work in an incremental fashion over the stream of distinct incoming IP addresses.
Figure~\ref{fig:anomDetectionOverview} shows a schematic overview of our algorithm which we summarize in the following steps.
\squishlist
\item (Step 1) Reduce the stream of incoming traffic to a stream of distinct source addresses.
\item (Step 2) Maintain a hash table of $N$ addresses. When a new distinct address arrives, enter it directly into the slot indicated by its hashed value and evict the address previously in this slot.
\item (Step 3) Incrementally update the required $Z(l, q)$ values for the prefixes of length $l$. Add one to the prefixes containing the new address and subtract one from prefixes containing the old (evicted) address.
\item (Step 4) Re-compute $\tilde{\tau}(q)$ and its estimated variance based on the updated $Z(l, q)$ values.
\item (Step 5) Compare the new estimate of $\tilde{\tau}(q)$ with the ``lagged'' estimate from $k$ distinct addresses ago using the two-sample Hotelling's $t^{2}$-test for multivariate normal distributions~\cite{hotelling}.\footnote{We assume independence between each $q$-value when translating the per-$q$ estimates to a multivariate normal.}
\item (Step 6) Take 1 minus the p-value of the $t^{2}$ test as the ``anomalous score''.
\squishend

The key idea of this procedure is to compare $\tilde{\tau}(q)$ estimated using the most recent address (or set of addresses) with a historic ``snapshot'' of the same estimator from several addresses previously.
In particular, step (5) tests the null hypothesis that the mean of this estimator is the same as $k$ addresses ago.
By the asymptotic consistency of $\tilde{\tau}(q)$, we are thus testing whether or not the new address is consistent with the same estimate of the previous $\tau(q)$.
Since the test rejects the null hypothesis if the p-value is less than some threshold, we flip the direction of the p-value to yield a metric or ``anomalous score'' that is larger for more anomalous (\ie less likely) addresses.

\begin{figure}[!htb]
    \vspace{-0.4cm}
    \centering
    \begin{subfigure}[t]{0.48\textwidth}
    \includegraphics[width=0.9\linewidth]{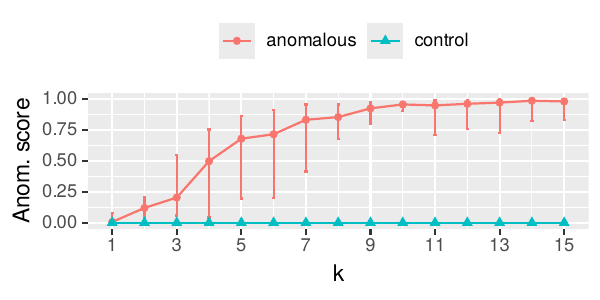}
    \vspace{-0.3cm}
    \caption{Effect of lag ($k$) on ``anomalous score''.}
    \label{fig:anomScore:lag}
    \end{subfigure}~
    \begin{subfigure}[t]{0.48\textwidth}
    \includegraphics[width=0.9\linewidth]{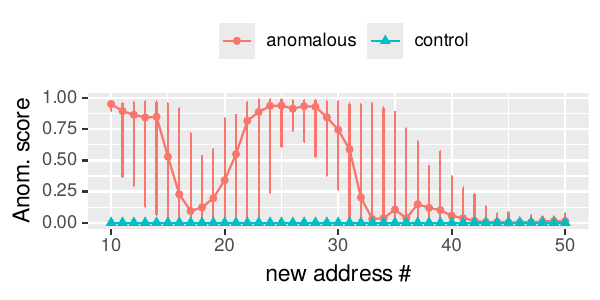}
    \vspace{-0.3cm}
    \caption{Evolution as addresses are added with $k = 10$.}
    \label{fig:anomScore:run}
    \end{subfigure}
    \vspace{-0.3cm}
    \caption{Impact of anomalous addresses compared to the original ``control'' addresses. Showing median, 5-th, and 95-th quantiles over 10 sampled distinct address sequences from CAIDA500k.}
    \label{fig:anomScore}
    \vspace{-0.4cm}
\end{figure}

To illustrate our approach, we consider a sample of $N$ = 50k distinct contiguously-observed addresses from the CAIDA500k dataset and two approaches to extending this set: {\em (i)} a ``control'' approach which simply takes the next addresses from CAIDA500k and {\em (ii)} an ``anomalous'' approach which generates the next addresses such that they fall in a previously-observed /8 prefix, but have their remaining 24 bits drawn uniformly at random.
Figure~\ref{fig:anomScore:lag} shows the effect of the lag parameter ($k$) on anomalous score for extenting addresses generated by both the ``anomalous'' and ``control'' approaches aggregated over ten independent samples from the CAIDA500k dataset.
We observe that even at $k = 2$, which amounts to testing the impact of just two addresses, the anomalous addresses cause a marked difference in anomalous score compared to the control addresses (in particular median of 0.12 compared to 0.00).
Longer lags increase this difference up to $k = 10$ where the anomalous addresses yield median anomalous score of 0.95 while the control address's score remains below $5 \cdot 10^{-12}$.
This indicates that our method is able to separate anomalous (\eg malicious) addresses from control (\eg benign) addresses for a variety of different lag values with high specificity and low false-positives.

To understand how the anomalous score evolves during normal operation as more distinct anomalous addresses replace previously-observed addresses in the hash table, we fix $k = 10$ and show in Figure~\ref{fig:anomScore:run} the anomalous score computed for each added address.
On the one hand, the maximum anomalous score produced for the ``control'' addresses over the 50 added addresses shown is $2.4\cdot 10^{-7}$ indicating the anomalous score does not increase significantly above zero for addresses following the normal ``benign'' structure.
On the other hand, the anomalous score for the anomalous addresses starts at a median of 0.95 and decreases as the anomalous addresses are gradually cycled in and the updated $\tilde{\tau}(q)$ moves away from the original structure. 
In Appendix~\ref{sec:appx-anomalousChanges} we confirm that the changes detected by this method are focused on negative moments (\ie scaling behavior of small prefixes) which is consistent with the observation that this approach is able to detect changes in scaling structure introduced by only a few (\eg two or three) anomalous addresses.

\noindent
\textbf{\em Summary.}
In this section we demonstrated the potential impact of multifractal analysis tools by introducing a novel approach for detecting structurally anomalous addresses. This approach augments present-day blocklist-based methods by providing a complementary ``anomalous score'' for the large number of addresses not found on any blocklists. Moreover, because it incrementally updates the structure function estimates, this approach can be efficiently deployed on real-time streams of network data.

\section{Discussion \& Outlook}
\label{sec:discussion}

As a new invariant of Internet traffic, the multifractal structure of observed IP addresses (together with its root cause resulting from the address allocation processes) is potentially transformative for several open networking problems. We next list a sampling of such problems but leave a more detailed treatment of them for future work. 

First, the online network traffic anomaly detection problem in \S~\ref{sec:implications} is a usecase example that directly motivates new questions of theoretical and practical interest such as {\em what types of security or performance events lead to changes in the observed address structure?} and {\em what is the cost for an attacker to infer the structure of observed addresses and control their traffic to match this structure?} Similarly, given the generative capabilities described in \S~\ref{sec:generative}, prior works on different aspects of network monitoring and management (\eg ~\cite{collins2007using,chen2007measuring,soldo2008filtering,kuvcera2020enabling}) can now be revived because evaluating their solutions’ performance in a principled manner as {\em a functions of} address structure is no longer stymied by data privacy concerns and the shortcomings of IP address datasets extracted from limited anonymized traffic traces. At the same time, the ability alluded to in \S~\ref{sec:evidence} to produce summary metrics that quantify IP address structure, but hide any IP address-specific information provides promising opportunities to boost the performance of new networking-related AI solutions by using standard behavioral features (\eg per IP address packet counts) in conjunction with structure-specific features (\eg prefixes which the address is a member of) for model training.

For a different set of problems, the starting point is our main finding that the multifractal address structure defines a new Internet invariant. 
One implication of this finding is that there is no ``typical'' prefix length that could be used for particular networking tasks. As a result, because they consider addresses at a fixed granularity, existing sketch-based systems for, say, traffic monitoring~\cite{sketchlearn,song2020fcm,compsensesketch,elasticsketch} inherently waste resources on less interesting (\eg sparse) regions of the address space. This in turn argues for alternative systems that use a variable prefix-length approach that is more resource-efficient because it allows focusing on more interesting (\eg dense) regions of the space (\eg~\cite{zapdos}). Another implication is that since multifractal scaling manifests in a pronounced ``cluster-within-cluster'' property, there will always be extremely sparse or entirely empty regions of the address space between these clusters, irrespective of the near-total allocation of the IPv4 address space. Finding ways to effectively exploit this property is especially important when developing algorithmic solutions intended to run on resource-limited systems (\eg hardware dataplane processor like a switch ASIC) that deal with IP addresses (\eg~\cite{dynatos,sonata,newton,zheng2022flymon}).

Lastly, our work points to two problem areas that are statistical in nature. First, despite a large literature on wavelet-based multifractal analysis that includes more traditional approaches that work directly with wavelet coefficients~\cite{arneodo2008wavelet,gilbert-etal99,gilbert-etal:2003} and more advanced approaches that consider wavelet-leaders~\cite{lashermes2005wavelet,jaffard-etal:2006,wendt2007multifractality,serrano2009wavelet}, we found using such approaches ``out-of-box” to be similarly ill-suited as the method-of-moments approach for dealing with the idiosyncrasies of discrete and finite IP spaces, especially their pronounced sparsity. This experience of ours calls for adapting these methods and the existing theory so we can have confidence in the inferences we make with the resulting estimators. A second related problem concerns estimating the multifractal spectrum $f(\alpha)$ and using this fine-grained digest of mulltifractal scaling behaviors in practice (\eg for fingerprinting the address structure observed at a particular vantage point in the network or at a particular point in time~\cite{kohler2002observed}). Although robust methods for estimating $f(\alpha)$ exist in the continuous case~\cite{wendt2007bootstrap}, their performance when applied in the context of discrete and finite IP spaces remain unknown and requires further studies.

\bibliographystyle{ACM-Reference-Format}
\bibliography{reference}


\begin{thebibliography}{76}

%%% the Web address.

\ifx \showCODEN    \undefined \def \showCODEN     #1{\unskip}     \fi
\ifx \showISBNx    \undefined \def \showISBNx     #1{\unskip}     \fi
\ifx \showISBNxiii \undefined \def \showISBNxiii  #1{\unskip}     \fi
\ifx \showISSN     \undefined \def \showISSN      #1{\unskip}     \fi
\ifx \showLCCN     \undefined \def \showLCCN      #1{\unskip}     \fi
\ifx \shownote     \undefined \def \shownote      #1{#1}          \fi
\ifx \showarticletitle \undefined \def \showarticletitle #1{#1}   \fi
\ifx \showURL      \undefined \def \showURL       {\relax}        \fi
% invisible to TeX
\providecommand\bibfield[2]{#2}
\providecommand\bibinfo[2]{#2}
\providecommand\natexlab[1]{#1}
\providecommand\showeprint[2][]{arXiv:#2}

\bibitem[rfc(1981a)]%
        {rfc790}
 \bibinfo{year}{1981}\natexlab{a}.
\newblock \bibinfo{title}{{Assigned numbers}}.
\newblock \bibinfo{howpublished}{RFC 790}.
\newblock
\href{https://doi.org/10.17487/RFC0790}{doi:\nolinkurl{10.17487/RFC0790}}


\bibitem[rfc(1981b)]%
        {rfc791}
 \bibinfo{year}{1981}\natexlab{b}.
\newblock \bibinfo{title}{{Internet Protocol}}.
\newblock \bibinfo{howpublished}{RFC 791}.
\newblock
\href{https://doi.org/10.17487/RFC0791}{doi:\nolinkurl{10.17487/RFC0791}}


\bibitem[AFRINIC({[n.\,d.]})]%
        {afrinicTransferPolicy}
\bibfield{author}{\bibinfo{person}{AFRINIC}.}
  \bibinfo{year}{[n.\,d.]}\natexlab{}.
\newblock \bibinfo{title}{Resource Transfers}.
\newblock
  \bibinfo{howpublished}{\url{https://afrinic.net/resources/transfers}}.
\newblock
\newblock
\shownote{Accessed: 2024-11}.


\bibitem[AFRINIC(2015)]%
        {afrinicInterProp}
\bibfield{author}{\bibinfo{person}{AFRINIC}.} \bibinfo{year}{2015}\natexlab{}.
\newblock \bibinfo{title}{Numer Resources Transfer Policy - {AFRINIC} -
  Regional Internet Registry for Africa}.
\newblock
  \bibinfo{howpublished}{\url{https://afrinic.net/policy/archive/number-resources-transfer-policy}}.
\newblock
\newblock
\shownote{Accessed: 2024-11}.


\bibitem[APNIC({[n.\,d.]})]%
        {spacebetween}
\bibfield{author}{\bibinfo{person}{APNIC}.}
  \bibinfo{year}{[n.\,d.]}\natexlab{}.
\newblock \bibinfo{title}{The space between IPv6 allocations: part 2}.
\newblock
  \bibinfo{howpublished}{\url{https://blog.apnic.net/2021/12/17/the-space-between-ipv6-allocations-part-2/}}.
\newblock
\newblock
\shownote{Accessed 2024-05-15}.


\bibitem[ARIN({[n.\,d.]})]%
        {arinTransfers}
\bibfield{author}{\bibinfo{person}{ARIN}.} \bibinfo{year}{[n.\,d.]}\natexlab{}.
\newblock \bibinfo{title}{Statistics: Specified Transfers of Internet Number
  Resources}.
\newblock
  \bibinfo{howpublished}{\url{https://account.arin.net/public/transfer-log/}}.
\newblock
\newblock
\shownote{Accessed: 2024-01}.


\bibitem[Arneodo et~al\mbox{.}(2008)]%
        {arneodo2008wavelet}
\bibfield{author}{\bibinfo{person}{Alain Arneodo}, \bibinfo{person}{Benjamin
  Audit}, \bibinfo{person}{Pierre Kestener}, {and} \bibinfo{person}{Stephane
  Roux}.} \bibinfo{year}{2008}\natexlab{}.
\newblock \showarticletitle{Wavelet-based multifractal analysis}.
\newblock \bibinfo{journal}{\emph{Scholarpedia}} \bibinfo{volume}{3},
  \bibinfo{number}{3} (\bibinfo{year}{2008}), \bibinfo{pages}{4103}.
\newblock


\bibitem[Avin et~al\mbox{.}(2020)]%
        {avin2020complexity}
\bibfield{author}{\bibinfo{person}{Chen Avin}, \bibinfo{person}{Manya Ghobadi},
  \bibinfo{person}{Chen Griner}, {and} \bibinfo{person}{Stefan Schmid}.}
  \bibinfo{year}{2020}\natexlab{}.
\newblock \showarticletitle{On the complexity of traffic traces and
  implications}.
\newblock \bibinfo{journal}{\emph{Proceedings of the ACM on Measurement and
  Analysis of Computing Systems}} \bibinfo{volume}{4}, \bibinfo{number}{1}
  (\bibinfo{year}{2020}), \bibinfo{pages}{1--29}.
\newblock


\bibitem[Barford et~al\mbox{.}(2006)]%
        {barford2006toward}
\bibfield{author}{\bibinfo{person}{Paul Barford}, \bibinfo{person}{Rob Nowak},
  \bibinfo{person}{Rebecca Willett}, {and} \bibinfo{person}{Vinod
  Yegneswaran}.} \bibinfo{year}{2006}\natexlab{}.
\newblock \showarticletitle{Toward a model for source addresses of internet
  background radiation}. In \bibinfo{booktitle}{\emph{Proc. of the Passive and
  Active Measurement Conference}}. Citeseer.
\newblock


\bibitem[CAIDA({[n.\,d.]})]%
        {caida2019}
\bibfield{author}{\bibinfo{person}{CAIDA}.}
  \bibinfo{year}{[n.\,d.]}\natexlab{}.
\newblock \bibinfo{title}{The {CAIDA UCSD} Anonymized {Internet} Traces Dataset
  - 2019}.
\newblock
  \bibinfo{howpublished}{\url{https://www.caida.org/data/monitors/passive-equinix-nyc.xml}}.
\newblock
\newblock
\shownote{Accessed: 2024-02}.


\bibitem[Chen and Ji(2007)]%
        {chen2007measuring}
\bibfield{author}{\bibinfo{person}{Zesheng Chen} {and} \bibinfo{person}{Chuanyi
  Ji}.} \bibinfo{year}{2007}\natexlab{}.
\newblock \showarticletitle{Measuring network-aware worm spreading ability}. In
  \bibinfo{booktitle}{\emph{IEEE INFOCOM 2007-26th IEEE International
  Conference on Computer Communications}}. IEEE, \bibinfo{pages}{116--124}.
\newblock


\bibitem[Collins et~al\mbox{.}(2007)]%
        {collins2007using}
\bibfield{author}{\bibinfo{person}{M~Patrick Collins},
  \bibinfo{person}{Timothy~J Shimeall}, \bibinfo{person}{Sidney Faber},
  \bibinfo{person}{Jeff Janies}, \bibinfo{person}{Rhiannon Weaver},
  \bibinfo{person}{Markus De~Shon}, {and} \bibinfo{person}{Joseph Kadane}.}
  \bibinfo{year}{2007}\natexlab{}.
\newblock \showarticletitle{Using uncleanliness to predict future botnet
  addresses}. In \bibinfo{booktitle}{\emph{Proceedings of the 7th ACM SIGCOMM
  conference on Internet measurement}}. \bibinfo{pages}{93--104}.
\newblock


\bibitem[Evertsz and Mandelbrot(1992)]%
        {evertsz1992multifractal}
\bibfield{author}{\bibinfo{person}{Carl~JG Evertsz} {and}
  \bibinfo{person}{Benoit~B Mandelbrot}.} \bibinfo{year}{1992}\natexlab{}.
\newblock \showarticletitle{Multifractal measures}.
\newblock \bibinfo{journal}{\emph{Chaos and fractals}}  \bibinfo{volume}{1992}
  (\bibinfo{year}{1992}), \bibinfo{pages}{921--953}.
\newblock


\bibitem[Fan et~al\mbox{.}({[n.\,d.]})]%
        {cryptopan}
\bibfield{author}{\bibinfo{person}{Jinliang Fan}, \bibinfo{person}{Jun Xu},
  \bibinfo{person}{Mostafa~H. Ammar}, {and} \bibinfo{person}{Sue Moon}.}
  \bibinfo{year}{[n.\,d.]}\natexlab{}.
\newblock \bibinfo{title}{Cryptopan}.
\newblock
  \bibinfo{howpublished}{\url{https://web.archive.org/web/20190430122100/https://www.cc.gatech.edu/computing/Networking/projects/cryptopan/}}.
\newblock
\newblock
\shownote{Accessed: 2024-02}.


\bibitem[Feal et~al\mbox{.}(2021)]%
        {feal2021blocklist}
\bibfield{author}{\bibinfo{person}{{\'A}lvaro Feal}, \bibinfo{person}{Pelayo
  Vallina}, \bibinfo{person}{Julien Gamba}, \bibinfo{person}{Sergio Pastrana},
  \bibinfo{person}{Antonio Nappa}, \bibinfo{person}{Oliver Hohlfeld},
  \bibinfo{person}{Narseo Vallina-Rodriguez}, {and} \bibinfo{person}{Juan
  Tapiador}.} \bibinfo{year}{2021}\natexlab{}.
\newblock \showarticletitle{Blocklist babel: On the transparency and dynamics
  of open source blocklisting}.
\newblock \bibinfo{journal}{\emph{IEEE Transactions on Network and Service
  Management}} \bibinfo{volume}{18}, \bibinfo{number}{2}
  (\bibinfo{year}{2021}), \bibinfo{pages}{1334--1349}.
\newblock


\bibitem[Feldmann et~al\mbox{.}(1998)]%
        {feldmann-sigcomm98}
\bibfield{author}{\bibinfo{person}{Anja Feldmann}, \bibinfo{person}{Anna~C
  Gilbert}, {and} \bibinfo{person}{Walter Willinger}.}
  \bibinfo{year}{1998}\natexlab{}.
\newblock \showarticletitle{Data networks as cascades: Investigating the
  multifractal nature of Internet WAN traffic}.
\newblock \bibinfo{journal}{\emph{ACM SIGCOMM Computer Communication Review}}
  \bibinfo{volume}{28}, \bibinfo{number}{4} (\bibinfo{year}{1998}),
  \bibinfo{pages}{42--55}.
\newblock


\bibitem[Floyd and Paxson(2001)]%
        {paxson-floyd:ton2001}
\bibfield{author}{\bibinfo{person}{Sally Floyd} {and} \bibinfo{person}{Vern
  Paxson}.} \bibinfo{year}{2001}\natexlab{}.
\newblock \showarticletitle{Difficulties in simulating the Internet}.
\newblock \bibinfo{journal}{\emph{IEEE/ACM Transactions on networking}}
  \bibinfo{volume}{9}, \bibinfo{number}{4} (\bibinfo{year}{2001}),
  \bibinfo{pages}{392--403}.
\newblock


\bibitem[Fontugne et~al\mbox{.}(2010)]%
        {mawilab}
\bibfield{author}{\bibinfo{person}{Romain Fontugne}, \bibinfo{person}{Pierre
  Borgnat}, \bibinfo{person}{Patrice Abry}, {and} \bibinfo{person}{Kensuke
  Fukuda}.} \bibinfo{year}{2010}\natexlab{}.
\newblock \showarticletitle{{MAWILab}: Combining Diverse Anomaly Detectors for
  Automated Anomaly Labeling and Performance Benchmarking}. In
  \bibinfo{booktitle}{\emph{Proceedings of the ACM Conference on emerging
  Networking EXperiments and Technologies (CoNEXT)}}.
  \bibinfo{numpages}{12}~pages.
\newblock


\bibitem[Fuller et~al\mbox{.}(1993)]%
        {rfc1519}
\bibfield{author}{\bibinfo{person}{Vince Fuller}, \bibinfo{person}{Tony Li},
  \bibinfo{person}{Kannan Varadhan}, {and} \bibinfo{person}{Jessica Yu}.}
  \bibinfo{year}{1993}\natexlab{}.
\newblock \bibinfo{title}{{Classless Inter-Domain Routing (CIDR): an Address
  Assignment and Aggregation Strategy}}.
\newblock \bibinfo{howpublished}{RFC 1519}.
\newblock
\href{https://doi.org/10.17487/RFC1519}{doi:\nolinkurl{10.17487/RFC1519}}


\bibitem[Gerich(1993)]%
        {rfc1466}
\bibfield{author}{\bibinfo{person}{Elise~P. Gerich}.}
  \bibinfo{year}{1993}\natexlab{}.
\newblock \bibinfo{title}{{Guidelines for Management of IP Address Space}}.
\newblock \bibinfo{howpublished}{RFC 1466}.
\newblock
\href{https://doi.org/10.17487/RFC1466}{doi:\nolinkurl{10.17487/RFC1466}}


\bibitem[Gilbert et~al\mbox{.}(1999)]%
        {gilbert-etal99}
\bibfield{author}{\bibinfo{person}{Anna~C Gilbert}, \bibinfo{person}{Walter
  Willinger}, {and} \bibinfo{person}{Anja Feldmann}.}
  \bibinfo{year}{1999}\natexlab{}.
\newblock \showarticletitle{Scaling analysis of conservative cascades, with
  applications to network traffic}.
\newblock \bibinfo{journal}{\emph{IEEE Transactions on Information Theory}}
  \bibinfo{volume}{45}, \bibinfo{number}{3} (\bibinfo{year}{1999}),
  \bibinfo{pages}{971--991}.
\newblock


\bibitem[Gupta et~al\mbox{.}(2018)]%
        {sonata}
\bibfield{author}{\bibinfo{person}{Arpit Gupta}, \bibinfo{person}{Rob
  Harrison}, \bibinfo{person}{Marco Canini}, \bibinfo{person}{Nick Feamster},
  \bibinfo{person}{Jennifer Rexford}, {and} \bibinfo{person}{Walter
  Willinger}.} \bibinfo{year}{2018}\natexlab{}.
\newblock \showarticletitle{Sonata: Query-driven streaming network telemetry}.
  In \bibinfo{booktitle}{\emph{Proceedings of the conference of the ACM Special
  Interest Group on Data Communication (SIGCOMM)}}. \bibinfo{pages}{357--371}.
\newblock


\bibitem[Hanan et~al\mbox{.}(2000)]%
        {hananLeftsided}
\bibfield{author}{\bibinfo{person}{WG Hanan}, \bibinfo{person}{J Gough}, {and}
  \bibinfo{person}{DM Heffernan}.} \bibinfo{year}{2000}\natexlab{}.
\newblock \showarticletitle{Left-sided multifractality in a binary random
  multiplicative cascade}.
\newblock \bibinfo{journal}{\emph{Physical Review E}} \bibinfo{volume}{63},
  \bibinfo{number}{1} (\bibinfo{year}{2000}), \bibinfo{pages}{011109}.
\newblock


\bibitem[Holley and Waymire(1992)]%
        {holley1992multifractal}
\bibfield{author}{\bibinfo{person}{Richard Holley} {and}
  \bibinfo{person}{Edward~C Waymire}.} \bibinfo{year}{1992}\natexlab{}.
\newblock \showarticletitle{Multifractal dimensions and scaling exponents for
  strongly bounded random cascades}.
\newblock \bibinfo{journal}{\emph{The Annals of Applied Probability}}
  (\bibinfo{year}{1992}), \bibinfo{pages}{819--845}.
\newblock


\bibitem[Hsu et~al\mbox{.}(2023)]%
        {hsu2023fiat}
\bibfield{author}{\bibinfo{person}{Amanda Hsu}, \bibinfo{person}{Frank Li},
  {and} \bibinfo{person}{Paul Pearce}.} \bibinfo{year}{2023}\natexlab{}.
\newblock \showarticletitle{Fiat lux: Illuminating IPv6 apportionment with
  different datasets}.
\newblock \bibinfo{journal}{\emph{Proceedings of the ACM on Measurement and
  Analysis of Computing Systems}} \bibinfo{volume}{7}, \bibinfo{number}{1}
  (\bibinfo{year}{2023}), \bibinfo{pages}{1--24}.
\newblock


\bibitem[Huang et~al\mbox{.}(2018)]%
        {sketchlearn}
\bibfield{author}{\bibinfo{person}{Qun Huang}, \bibinfo{person}{Patrick~PC
  Lee}, {and} \bibinfo{person}{Yungang Bao}.} \bibinfo{year}{2018}\natexlab{}.
\newblock \showarticletitle{Sketchlearn: Relieving user burdens in approximate
  measurement with automated statistical inference}. In
  \bibinfo{booktitle}{\emph{Proceedings of the conference of the ACM Special
  Interest Group on Data Communication (SIGCOMM)}}. \bibinfo{pages}{576--590}.
\newblock


\bibitem[Huang et~al\mbox{.}(2021)]%
        {compsensesketch}
\bibfield{author}{\bibinfo{person}{Qun Huang}, \bibinfo{person}{Siyuan Sheng},
  \bibinfo{person}{Xiang Chen}, \bibinfo{person}{Yungang Bao},
  \bibinfo{person}{Rui Zhang}, \bibinfo{person}{Yanwei Xu}, {and}
  \bibinfo{person}{Gong Zhang}.} \bibinfo{year}{2021}\natexlab{}.
\newblock \showarticletitle{Toward Nearly-Zero-Error Sketching via Compressive
  Sensing.}. In \bibinfo{booktitle}{\emph{NSDI}}. \bibinfo{pages}{1027--1044}.
\newblock


\bibitem[IANA({[n.\,d.]})]%
        {ianaipv6}
\bibfield{author}{\bibinfo{person}{IANA}.} \bibinfo{year}{[n.\,d.]}\natexlab{}.
\newblock \bibinfo{title}{{IPv6} Global Unicast Address Assignments}.
\newblock
  \bibinfo{howpublished}{\url{https://www.iana.org/assignments/ipv6-unicast-address-assignments/ipv6-unicast-address-assignments.xhtml}}.
\newblock
\newblock
\shownote{Accessed: 2024-01}.


\bibitem[IANA(2023)]%
        {ianaipv4}
\bibfield{author}{\bibinfo{person}{IANA}.} \bibinfo{year}{2023}\natexlab{}.
\newblock \bibinfo{title}{{IPv4} Address Space Registry}.
\newblock
  \bibinfo{howpublished}{\url{https://www.iana.org/assignments/ipv4-address-space/ipv4-address-space.xhtml}}.
\newblock
\newblock
\shownote{Accessed: 2024-01}.


\bibitem[ICANN({[n.\,d.]})]%
        {ipv4exhaustion}
\bibfield{author}{\bibinfo{person}{ICANN}.}
  \bibinfo{year}{[n.\,d.]}\natexlab{}.
\newblock \bibinfo{title}{Available Pool of Unallocated {IPv4} Internet
  Addresses Now Completely Emptied}.
\newblock
  \bibinfo{howpublished}{\url{https://itp.cdn.icann.org/en/files/announcements/release-03feb11-en.pdf}}.
\newblock
\newblock
\shownote{Accessed: 2024-01}.


\bibitem[Jaffard et~al\mbox{.}(2007)]%
        {jaffard-etal:2006}
\bibfield{author}{\bibinfo{person}{St{\'e}phane Jaffard},
  \bibinfo{person}{Bruno Lashermes}, {and} \bibinfo{person}{Patrice Abry}.}
  \bibinfo{year}{2007}\natexlab{}.
\newblock \showarticletitle{Wavelet leaders in multifractal analysis}. In
  \bibinfo{booktitle}{\emph{Wavelet analysis and applications}}. Springer,
  \bibinfo{pages}{201--246}.
\newblock


\bibitem[Kogan and Corrigan-Gibbs(2021)]%
        {kogan2021private}
\bibfield{author}{\bibinfo{person}{Dmitry Kogan} {and} \bibinfo{person}{Henry
  Corrigan-Gibbs}.} \bibinfo{year}{2021}\natexlab{}.
\newblock \showarticletitle{Private blocklist lookups with checklist}. In
  \bibinfo{booktitle}{\emph{30th USENIX security symposium (USENIX Security
  21)}}. \bibinfo{pages}{875--892}.
\newblock


\bibitem[Kohler et~al\mbox{.}(2002)]%
        {kohler2002observed}
\bibfield{author}{\bibinfo{person}{Eddie Kohler}, \bibinfo{person}{Jinyang Li},
  \bibinfo{person}{Vern Paxson}, {and} \bibinfo{person}{Scott Shenker}.}
  \bibinfo{year}{2002}\natexlab{}.
\newblock \showarticletitle{Observed structure of addresses in IP traffic}. In
  \bibinfo{booktitle}{\emph{Proceedings of the 2nd ACM SIGCOMM Workshop on
  Internet measurment}}. \bibinfo{pages}{253--266}.
\newblock


\bibitem[Kohler et~al\mbox{.}(2006)]%
        {kohler2006observed}
\bibfield{author}{\bibinfo{person}{Eddie Kohler}, \bibinfo{person}{Jinyang Li},
  \bibinfo{person}{Vern Paxson}, {and} \bibinfo{person}{Scott Shenker}.}
  \bibinfo{year}{2006}\natexlab{}.
\newblock \showarticletitle{Observed structure of addresses in IP traffic}.
\newblock \bibinfo{journal}{\emph{IEEE/ACM Transactions on Networking}}
  \bibinfo{volume}{14}, \bibinfo{number}{6} (\bibinfo{year}{2006}),
  \bibinfo{pages}{1207--1218}.
\newblock


\bibitem[Ku{\v{c}}era et~al\mbox{.}(2020)]%
        {kuvcera2020enabling}
\bibfield{author}{\bibinfo{person}{Jan Ku{\v{c}}era},
  \bibinfo{person}{Diana~Andreea Popescu}, \bibinfo{person}{Han Wang},
  \bibinfo{person}{Andrew Moore}, \bibinfo{person}{Jan Ko{\v{r}}enek}, {and}
  \bibinfo{person}{Gianni Antichi}.} \bibinfo{year}{2020}\natexlab{}.
\newblock \showarticletitle{Enabling event-triggered data plane monitoring}. In
  \bibinfo{booktitle}{\emph{Proceedings of the Symposium on SDN Research}}.
  \bibinfo{pages}{14--26}.
\newblock


\bibitem[Lashermes et~al\mbox{.}(2004)]%
        {lashermes2004new}
\bibfield{author}{\bibinfo{person}{Bruno Lashermes}, \bibinfo{person}{Patrice
  Abry}, {and} \bibinfo{person}{Pierre Chainais}.}
  \bibinfo{year}{2004}\natexlab{}.
\newblock \showarticletitle{New insights into the estimation of scaling
  exponents}.
\newblock \bibinfo{journal}{\emph{International Journal of Wavelets,
  Multiresolution and Information Processing}} \bibinfo{volume}{2},
  \bibinfo{number}{04} (\bibinfo{year}{2004}), \bibinfo{pages}{497--523}.
\newblock


\bibitem[Lashermes et~al\mbox{.}(2005)]%
        {lashermes2005wavelet}
\bibfield{author}{\bibinfo{person}{Bruno Lashermes},
  \bibinfo{person}{St{\'e}phane Jaffard}, {and} \bibinfo{person}{Patrice
  Abry}.} \bibinfo{year}{2005}\natexlab{}.
\newblock \showarticletitle{Wavelet leader based multifractal analysis}. In
  \bibinfo{booktitle}{\emph{Proceedings.(ICASSP'05). IEEE International
  Conference on Acoustics, Speech, and Signal Processing, 2005.}},
  Vol.~\bibinfo{volume}{4}. IEEE, \bibinfo{pages}{iv--161}.
\newblock


\bibitem[L{\'e}vy~V{\'e}hel and Riedi(1997)]%
        {riedi-lv:1997}
\bibfield{author}{\bibinfo{person}{Jacques L{\'e}vy~V{\'e}hel} {and}
  \bibinfo{person}{Rudolf Riedi}.} \bibinfo{year}{1997}\natexlab{}.
\newblock \showarticletitle{Fractional Brownian motion and data traffic
  modeling: The other end of the spectrum}. In
  \bibinfo{booktitle}{\emph{Fractals in Engineering: from theory to industrial
  applications}}. Springer, \bibinfo{pages}{185--202}.
\newblock


\bibitem[Li et~al\mbox{.}(2021)]%
        {li2021clairvoyance}
\bibfield{author}{\bibinfo{person}{Vector~Guo Li}, \bibinfo{person}{Gautam
  Akiwate}, \bibinfo{person}{Kirill Levchenko}, \bibinfo{person}{Geoffrey~M
  Voelker}, {and} \bibinfo{person}{Stefan Savage}.}
  \bibinfo{year}{2021}\natexlab{}.
\newblock \showarticletitle{Clairvoyance: Inferring blocklist use on the
  internet}. In \bibinfo{booktitle}{\emph{Passive and Active Measurement: 22nd
  International Conference, PAM 2021, Virtual Event, March 29--April 1, 2021,
  Proceedings 22}}. Springer, \bibinfo{pages}{57--75}.
\newblock


\bibitem[Mandelbrot.(1990)]%
        {mandelbrot:1990}
\bibfield{author}{\bibinfo{person}{B.~B. Mandelbrot.}}
  \bibinfo{year}{1990.}\natexlab{}.
\newblock \showarticletitle{Limit lognormal multifractal measures}.
\newblock \bibinfo{journal}{\emph{Frontiers of Physics: Proceedings of the
  Landau Memorial Conference}} (\bibinfo{year}{1990.}),
  \bibinfo{pages}{309–--340}.
\newblock


\bibitem[Mandelbrot(1990)]%
        {mandelbrotAnomalous}
\bibfield{author}{\bibinfo{person}{Benoit~B Mandelbrot}.}
  \bibinfo{year}{1990}\natexlab{}.
\newblock \showarticletitle{New “anomalous” multiplicative multifractals:
  Left sided ƒ ($\alpha$) and the modelling of DLA}.
\newblock \bibinfo{journal}{\emph{Physica A: Statistical Mechanics and its
  Applications}} \bibinfo{volume}{168}, \bibinfo{number}{1}
  (\bibinfo{year}{1990}), \bibinfo{pages}{95--111}.
\newblock


\bibitem[Mannersalo and Norros(1997)]%
        {mannersalo-norros:1997}
\bibfield{author}{\bibinfo{person}{Petteri Mannersalo} {and}
  \bibinfo{person}{Ilkka Norros}.} \bibinfo{year}{1997}\natexlab{}.
\newblock \bibinfo{title}{Multifractal analysis of real ATM traffic: a first
  look}.
\newblock


\bibitem[Minshall({[n.\,d.]})]%
        {tcpdpriv}
\bibfield{author}{\bibinfo{person}{Greg Minshall}.}
  \bibinfo{year}{[n.\,d.]}\natexlab{}.
\newblock \bibinfo{title}{Program for Eliminating Confidential Information from
  Traces}.
\newblock
  \bibinfo{howpublished}{\url{https://fly.isti.cnr.it/software/tcpdpriv/}}.
\newblock
\newblock
\shownote{Accessed: 2024-02}.


\bibitem[Misa et~al\mbox{.}(2024)]%
        {zapdos}
\bibfield{author}{\bibinfo{person}{Chris Misa}, \bibinfo{person}{Ramakrishnan
  Durairajan}, \bibinfo{person}{Reza Rejaie}, {and} \bibinfo{person}{Walter
  Willinger}.} \bibinfo{year}{2024}\natexlab{}.
\newblock \showarticletitle{Leveraging Prefix Structure to Detect Volumetric
  DDoS Attack Signatures with Programmable Switches}.
\newblock \bibinfo{journal}{\emph{IEEE Symposium on Security and Privacy (S and
  P) (Oakland)}} (\bibinfo{year}{2024}).
\newblock


\bibitem[Misa et~al\mbox{.}(2022)]%
        {dynatos}
\bibfield{author}{\bibinfo{person}{Chris Misa}, \bibinfo{person}{Walt
  O'Connor}, \bibinfo{person}{Ramakrishnan Durairajan}, \bibinfo{person}{Reza
  Rejaie}, {and} \bibinfo{person}{Walter Willinger}.}
  \bibinfo{year}{2022}\natexlab{}.
\newblock \showarticletitle{Dynamic Scheduling of Approximate Telemetry
  Queries}. In \bibinfo{booktitle}{\emph{19th USENIX Symposium on Networked
  Systems Design and Implementation (NSDI 22)}}. \bibinfo{pages}{701--717}.
\newblock


\bibitem[Nordhausen({[n.\,d.]})]%
        {hotelling}
\bibfield{author}{\bibinfo{person}{Klaus Nordhausen}.}
  \bibinfo{year}{[n.\,d.]}\natexlab{}.
\newblock \bibinfo{title}{R: {Hotelling's} T2 Test}.
\newblock
  \bibinfo{howpublished}{\url{https://search.r-project.org/CRAN/refmans/DescTools/html/HotellingsT.html}}.
\newblock
\newblock
\shownote{Accessed: 2025-01}.


\bibitem[Nowak(1998)]%
        {nowak:1998}
\bibfield{author}{\bibinfo{person}{Robert~D Nowak}.}
  \bibinfo{year}{1998}\natexlab{}.
\newblock \showarticletitle{Fractal modeling and analysis of poisson
  processes}. In \bibinfo{booktitle}{\emph{Conference Record of Thirty-Second
  Asilomar Conference on Signals, Systems and Computers (Cat. No. 98CH36284)}},
  Vol.~\bibinfo{volume}{1}. IEEE, \bibinfo{pages}{727--731}.
\newblock


\bibitem[Olsen(1995)]%
        {olsen1995multifractal}
\bibfield{author}{\bibinfo{person}{Lars Olsen}.}
  \bibinfo{year}{1995}\natexlab{}.
\newblock \showarticletitle{A multifractal formalism}.
\newblock \bibinfo{journal}{\emph{Advances in mathematics}}
  \bibinfo{volume}{116}, \bibinfo{number}{1} (\bibinfo{year}{1995}),
  \bibinfo{pages}{82--196}.
\newblock


\bibitem[Ossiander and Waymire(2000)]%
        {ossiander2000statistical}
\bibfield{author}{\bibinfo{person}{Mina Ossiander} {and}
  \bibinfo{person}{Edward~C Waymire}.} \bibinfo{year}{2000}\natexlab{}.
\newblock \showarticletitle{Statistical estimation for multiplicative
  cascades}.
\newblock \bibinfo{journal}{\emph{The Annals of Statistics}}
  \bibinfo{volume}{28}, \bibinfo{number}{6} (\bibinfo{year}{2000}),
  \bibinfo{pages}{1533--1560}.
\newblock


\bibitem[Proofpoint({[n.\,d.]})]%
        {proofpoint}
\bibfield{author}{\bibinfo{person}{Inc. Proofpoint}.}
  \bibinfo{year}{[n.\,d.]}\natexlab{}.
\newblock \bibinfo{title}{Proofpoint Emerging Threats Rules}.
\newblock \bibinfo{howpublished}{\url{https://rules.emergingthreats.net/}}.
\newblock
\newblock
\shownote{Accessed: 2025-01}.


\bibitem[Ramanathan et~al\mbox{.}(2020)]%
        {ramanathan2020quantifying}
\bibfield{author}{\bibinfo{person}{Sivaramakrishnan Ramanathan},
  \bibinfo{person}{Anushah Hossain}, \bibinfo{person}{Jelena Mirkovic},
  \bibinfo{person}{Minlan Yu}, {and} \bibinfo{person}{Sadia Afroz}.}
  \bibinfo{year}{2020}\natexlab{}.
\newblock \showarticletitle{Quantifying the impact of blocklisting in the age
  of address reuse}. In \bibinfo{booktitle}{\emph{Proceedings of the ACM
  Internet Measurement Conference}}. \bibinfo{pages}{360--369}.
\newblock


\bibitem[Resnick et~al\mbox{.}(2003a)]%
        {gilbert-etal:2003}
\bibfield{author}{\bibinfo{person}{Sidney Resnick}, \bibinfo{person}{Gennady
  Samorodnitsky}, \bibinfo{person}{Anna Gilbert}, {and} \bibinfo{person}{Walter
  Willinger}.} \bibinfo{year}{2003}\natexlab{a}.
\newblock \showarticletitle{Wavelet analysis of conservative cascades}.
\newblock \bibinfo{journal}{\emph{Bernoulli}} \bibinfo{volume}{9},
  \bibinfo{number}{1} (\bibinfo{year}{2003}), \bibinfo{pages}{97--135}.
\newblock


\bibitem[Resnick et~al\mbox{.}(2003b)]%
        {resnick2003wavelet}
\bibfield{author}{\bibinfo{person}{Sidney Resnick}, \bibinfo{person}{Gennady
  Samorodnitsky}, \bibinfo{person}{Anna Gilbert}, {and} \bibinfo{person}{Walter
  Willinger}.} \bibinfo{year}{2003}\natexlab{b}.
\newblock \showarticletitle{Wavelet analysis of conservative cascades}.
\newblock \bibinfo{journal}{\emph{Bernoulli}} \bibinfo{volume}{9},
  \bibinfo{number}{1} (\bibinfo{year}{2003}), \bibinfo{pages}{97--135}.
\newblock


\bibitem[Riedi(1995)]%
        {riedi1995improved}
\bibfield{author}{\bibinfo{person}{Rolf Riedi}.}
  \bibinfo{year}{1995}\natexlab{}.
\newblock \showarticletitle{An improved multifractal formalism and self-similar
  measures}.
\newblock \bibinfo{journal}{\emph{J. Math. Anal. Appl.}} \bibinfo{volume}{189},
  \bibinfo{number}{2} (\bibinfo{year}{1995}), \bibinfo{pages}{462--490}.
\newblock


\bibitem[Riedi et~al\mbox{.}(1999)]%
        {riedi-etal99}
\bibfield{author}{\bibinfo{person}{Rudolf~H Riedi}, \bibinfo{person}{Matthew~S
  Crouse}, \bibinfo{person}{Vinay~J Ribeiro}, {and} \bibinfo{person}{Richard~G
  Baraniuk}.} \bibinfo{year}{1999}\natexlab{}.
\newblock \showarticletitle{A multifractal wavelet model with application to
  network traffic}.
\newblock \bibinfo{journal}{\emph{IEEE transactions on Information Theory}}
  \bibinfo{volume}{45}, \bibinfo{number}{3} (\bibinfo{year}{1999}),
  \bibinfo{pages}{992--1018}.
\newblock


\bibitem[RIPE({[n.\,d.]})]%
        {ripeIPv6NewPolicy}
\bibfield{author}{\bibinfo{person}{RIPE}.} \bibinfo{year}{[n.\,d.]}\natexlab{}.
\newblock \bibinfo{title}{Updated Approach to IPv6 Transfer Requests}.
\newblock
  \bibinfo{howpublished}{\url{https://www.ripe.net/about-us/news/updated-approach-to-ipv6-transfer-requests/}}.
\newblock
\newblock
\shownote{Accessed: 2024-11}.


\bibitem[RIPE(2012)]%
        {ripeAllocVsAssign}
\bibfield{author}{\bibinfo{person}{RIPE}.} \bibinfo{year}{2012}\natexlab{}.
\newblock \bibinfo{title}{{IPv6} Address Allocation and Assignment Policy}.
\newblock
  \bibinfo{howpublished}{\url{https://www.ripe.net/publications/docs/ripe-552/}}.
\newblock
\newblock
\shownote{Accessed: 2024-10}.


\bibitem[Salat et~al\mbox{.}(2017)]%
        {salat2017multifractal}
\bibfield{author}{\bibinfo{person}{Hadrien Salat}, \bibinfo{person}{Roberto
  Murcio}, {and} \bibinfo{person}{Elsa Arcaute}.}
  \bibinfo{year}{2017}\natexlab{}.
\newblock \showarticletitle{Multifractal methodology}.
\newblock \bibinfo{journal}{\emph{Physica A: Statistical Mechanics and its
  Applications}}  \bibinfo{volume}{473} (\bibinfo{year}{2017}),
  \bibinfo{pages}{467--487}.
\newblock


\bibitem[Schmidt and Wilhelm({[n.\,d.]})]%
        {ripeIPv6Stockpiling}
\bibfield{author}{\bibinfo{person}{Marco Schmidt} {and} \bibinfo{person}{Rene
  Wilhelm}.} \bibinfo{year}{[n.\,d.]}\natexlab{}.
\newblock \bibinfo{title}{{IPv6} Stockpililng: A Trojan Horse in Our Midst?}
\newblock
  \bibinfo{howpublished}{\url{https://labs.ripe.net/author/marco_schmidt/ipv6-stockpiling-a-trojan-horse-in-our-midst/}}.
\newblock
\newblock
\shownote{Accessed: 2024-11}.


\bibitem[Serrano and Figliola(2009)]%
        {serrano2009wavelet}
\bibfield{author}{\bibinfo{person}{Eduardo Serrano} {and}
  \bibinfo{person}{Alejandra Figliola}.} \bibinfo{year}{2009}\natexlab{}.
\newblock \showarticletitle{Wavelet leaders: a new method to estimate the
  multifractal singularity spectra}.
\newblock \bibinfo{journal}{\emph{Physica A: Statistical Mechanics and its
  Applications}} \bibinfo{volume}{388}, \bibinfo{number}{14}
  (\bibinfo{year}{2009}), \bibinfo{pages}{2793--2805}.
\newblock


\bibitem[Soldo et~al\mbox{.}(2008)]%
        {soldo2008filtering}
\bibfield{author}{\bibinfo{person}{Fabio Soldo}, \bibinfo{person}{Karim
  El~Defrawy}, \bibinfo{person}{Athina Markopoulou},
  \bibinfo{person}{Balachander Krishnamurthy}, {and} \bibinfo{person}{Jacobus
  van~der Merwe}.} \bibinfo{year}{2008}\natexlab{}.
\newblock \showarticletitle{Filtering sources of unwanted traffic}. In
  \bibinfo{booktitle}{\emph{2008 Information Theory and Applications
  Workshop}}. IEEE, \bibinfo{pages}{199--208}.
\newblock


\bibitem[Sommers and Raffensperger(2012)]%
        {sommers2012efficient}
\bibfield{author}{\bibinfo{person}{Joel Sommers} {and} \bibinfo{person}{John
  Raffensperger}.} \bibinfo{year}{2012}\natexlab{}.
\newblock \showarticletitle{Efficient and realistic generation of IP
  addresses}. In \bibinfo{booktitle}{\emph{4th International ICST Conference on
  Simulation Tools and Techniques}}.
\newblock


\bibitem[Song et~al\mbox{.}(2020)]%
        {song2020fcm}
\bibfield{author}{\bibinfo{person}{Cha~Hwan Song},
  \bibinfo{person}{Pravein~Govindan Kannan}, \bibinfo{person}{Bryan Kian~Hsiang
  Low}, {and} \bibinfo{person}{Mun~Choon Chan}.}
  \bibinfo{year}{2020}\natexlab{}.
\newblock \showarticletitle{Fcm-sketch: generic network measurements with data
  plane support}. In \bibinfo{booktitle}{\emph{Proceedings of the 16th
  International Conference on emerging Networking EXperiments and
  Technologies}}. \bibinfo{pages}{78--92}.
\newblock


\bibitem[Stampar({[n.\,d.]})]%
        {ipsum}
\bibfield{author}{\bibinfo{person}{Miroslav Stampar}.}
  \bibinfo{year}{[n.\,d.]}\natexlab{}.
\newblock \bibinfo{title}{ipsum: Daily feed of bad IPs (with blacklist hit
  scores)}.
\newblock \bibinfo{howpublished}{\url{https://github.com/stamparm/ipsum}}.
\newblock
\newblock
\shownote{Accessed: 2025-01}.


\bibitem[Trading({[n.\,d.]})]%
        {afrinicTransferTrading}
\bibfield{author}{\bibinfo{person}{IP Trading}.}
  \bibinfo{year}{[n.\,d.]}\natexlab{}.
\newblock \bibinfo{title}{{AFRINIC IP} Transfer Process}.
\newblock
  \bibinfo{howpublished}{\url{https://iptrading.com/afrinic-transfer-process/}}.
\newblock
\newblock
\shownote{Accessed: 2024-11}.


\bibitem[Tsaousis({[n.\,d.]})]%
        {firehol}
\bibfield{author}{\bibinfo{person}{Costa Tsaousis}.}
  \bibinfo{year}{[n.\,d.]}\natexlab{}.
\newblock \bibinfo{title}{{FireHOL} {IP} Lists | {IP} Blacklist | {IP}
  Blocklists | {IP} Reputation}.
\newblock \bibinfo{howpublished}{\url{http://iplists.firehol.org/}}.
\newblock
\newblock
\shownote{Accessed: 2025-01}.


\bibitem[Vegoda({[n.\,d.]})]%
        {otherAllocVsAssign}
\bibfield{author}{\bibinfo{person}{Leo Vegoda}.}
  \bibinfo{year}{[n.\,d.]}\natexlab{}.
\newblock \bibinfo{title}{Assignments, Allocations and Temporary Transfers}.
\newblock \bibinfo{howpublished}{https://ipv4.global/events/assignments/}.
\newblock
\newblock
\shownote{Accessed: 2024-10}.


\bibitem[Veitch et~al\mbox{.}(2005)]%
        {veitch-etal05}
\bibfield{author}{\bibinfo{person}{Darryl Veitch}, \bibinfo{person}{Nicolas
  Hohn}, {and} \bibinfo{person}{Patrice Abry}.}
  \bibinfo{year}{2005}\natexlab{}.
\newblock \showarticletitle{Multifractality in TCP/IP traffic: the case
  against}.
\newblock \bibinfo{journal}{\emph{Computer Networks}} \bibinfo{volume}{48},
  \bibinfo{number}{3} (\bibinfo{year}{2005}), \bibinfo{pages}{293--313}.
\newblock


\bibitem[Wendt and Abry(2007a)]%
        {wendt-abry:2007}
\bibfield{author}{\bibinfo{person}{Herwig Wendt} {and} \bibinfo{person}{Patrice
  Abry}.} \bibinfo{year}{2007}\natexlab{a}.
\newblock \showarticletitle{Multifractality tests using bootstrapped wavelet
  leaders}.
\newblock \bibinfo{journal}{\emph{IEEE Transactions on Signal Processing}}
  \bibinfo{volume}{55}, \bibinfo{number}{10} (\bibinfo{year}{2007}),
  \bibinfo{pages}{4811--4820}.
\newblock


\bibitem[Wendt and Abry(2007b)]%
        {wendt2007multifractality}
\bibfield{author}{\bibinfo{person}{Herwig Wendt} {and} \bibinfo{person}{Patrice
  Abry}.} \bibinfo{year}{2007}\natexlab{b}.
\newblock \showarticletitle{Multifractality tests using bootstrapped wavelet
  leaders}.
\newblock \bibinfo{journal}{\emph{IEEE Transactions on Signal Processing}}
  \bibinfo{volume}{55}, \bibinfo{number}{10} (\bibinfo{year}{2007}),
  \bibinfo{pages}{4811--4820}.
\newblock


\bibitem[Wendt et~al\mbox{.}(2007)]%
        {wendt2007bootstrap}
\bibfield{author}{\bibinfo{person}{Herwig Wendt}, \bibinfo{person}{Patrice
  Abry}, {and} \bibinfo{person}{Stephine Jaffard}.}
  \bibinfo{year}{2007}\natexlab{}.
\newblock \showarticletitle{Bootstrap for empirical multifractal analysis}.
\newblock \bibinfo{journal}{\emph{IEEE signal processing magazine}}
  \bibinfo{volume}{24}, \bibinfo{number}{4} (\bibinfo{year}{2007}),
  \bibinfo{pages}{38--48}.
\newblock


\bibitem[Wikipedia({[n.\,d.]})]%
        {afiliasWiki}
\bibfield{author}{\bibinfo{person}{Wikipedia}.}
  \bibinfo{year}{[n.\,d.]}\natexlab{}.
\newblock \bibinfo{title}{Afilias}.
\newblock \bibinfo{howpublished}{https://en.wikipedia.org/wiki/Afilias}.
\newblock
\newblock
\shownote{Accessed: 2024-10}.


\bibitem[www.blocklist.de({[n.\,d.]})]%
        {blocklistde}
\bibfield{author}{\bibinfo{person}{www.blocklist.de}.}
  \bibinfo{year}{[n.\,d.]}\natexlab{}.
\newblock \bibinfo{title}{Fail2Ban Reporting Service}.
\newblock \bibinfo{howpublished}{\url{https://www.blocklist.de/en/index.html}}.
\newblock
\newblock
\shownote{Accessed: 2025-01}.


\bibitem[Yang et~al\mbox{.}(2018)]%
        {elasticsketch}
\bibfield{author}{\bibinfo{person}{Tong Yang}, \bibinfo{person}{Jie Jiang},
  \bibinfo{person}{Peng Liu}, \bibinfo{person}{Qun Huang},
  \bibinfo{person}{Junzhi Gong}, \bibinfo{person}{Yang Zhou},
  \bibinfo{person}{Rui Miao}, \bibinfo{person}{Xiaoming Li}, {and}
  \bibinfo{person}{Steve Uhlig}.} \bibinfo{year}{2018}\natexlab{}.
\newblock \showarticletitle{Elastic sketch: {Adaptive} and fast network-wide
  measurements}. In \bibinfo{booktitle}{\emph{Proceedings of the conference of
  the ACM Special Interest Group on Data Communication (SIGCOMM)}}. ACM,
  \bibinfo{pages}{561--575}.
\newblock


\bibitem[Zheng et~al\mbox{.}(2022)]%
        {zheng2022flymon}
\bibfield{author}{\bibinfo{person}{Hao Zheng}, \bibinfo{person}{Chen Tian},
  \bibinfo{person}{Tong Yang}, \bibinfo{person}{Huiping Lin},
  \bibinfo{person}{Chang Liu}, \bibinfo{person}{Zhaochen Zhang},
  \bibinfo{person}{Wanchun Dou}, {and} \bibinfo{person}{Guihai Chen}.}
  \bibinfo{year}{2022}\natexlab{}.
\newblock \showarticletitle{Flymon: enabling on-the-fly task reconfiguration
  for network measurement}. In \bibinfo{booktitle}{\emph{Proceedings of the ACM
  SIGCOMM 2022 Conference}}. \bibinfo{pages}{486--502}.
\newblock


\bibitem[Zhou et~al\mbox{.}(2020)]%
        {newton}
\bibfield{author}{\bibinfo{person}{Yu Zhou}, \bibinfo{person}{Dai Zhang},
  \bibinfo{person}{Kai Gao}, \bibinfo{person}{Chen Sun},
  \bibinfo{person}{Jiamin Cao}, \bibinfo{person}{Yangyang Wang},
  \bibinfo{person}{Mingwei Xu}, {and} \bibinfo{person}{Jianping Wu}.}
  \bibinfo{year}{2020}\natexlab{}.
\newblock \showarticletitle{Newton: {Intent-driven} network traffic
  monitoring}. In \bibinfo{booktitle}{\emph{Proceedings of the ACM Conference
  on emerging Networking EXperiments and Technologies (CoNEXT)}}.
  \bibinfo{pages}{295--308}.
\newblock


\end{thebibliography}

\appendix

\section{Appendix: Mathematical Background}
\label{sec:appx-mathBackground}

\subsection{Multifractal analysis in a nutshell}
\label{ssec:multifracatal-analysis}

For a mathematical treatment of multifractals (see, for example ~\cite{evertsz1992multifractal,salat2017multifractal} and references therein), it is common to study the irregularities or singularities of (finite) measures $\mu$ defined on some subset $E$ of the  1D or 2D space in terms of a pointwise scaling exponent called the \textbf{H\"{o}lder exponent} $\alpha(x)$. This exponent quantifies the degree of local regularity of $\mu$ at the point $x \in E$ and satisfies 
\begin{equation}
    \mu(B(r,x)) \approx r^{\alpha(x)} ~~~\text{as}~~ r \to 0, 
\end{equation}
where $B(r,x)$ denotes a ball of radius $r>0$ around $x \in E$. The measure $\mu$ is called {\bf multifractal} if this H\"{o}lder exponent varies as $x$ varies; $\mu$ is {\bf monofractal} if a single H\"{o}lder exponent $\alpha_0$ describes the singularities of $\mu$~\cite{evertsz1992multifractal}. {\bf Multifractal analysis} is concerned with effectively summarizing the complex and seemingly erratic nature in which the H\"{o}lder exponent can fluctuate for different values of $x \in E$ and providing practitioners with a succinct global description of these fluctuations. 

One such description is geometrical in nature and is based on the so-called \textbf{histogram method}. Informally, this method uses the concept of \textbf{fractal dimension} to quantify each iso-scaling set $E_\alpha = \{ x \in E, x \mbox{~has H\"{o}lder exponent~}\alpha \}$ and then computes the \textbf{multifractal spectrum} $f(\alpha)$, where for each $\alpha$, $f(\alpha)$ denotes the fractal dimension of the iso-scaling set $E_\alpha$. The multifractal spectrum thus encapsulates the properties of the different iso-scaling sets and can therefore be understood as providing a parsimonious quantitative digest and characterization of the complex fine-grained scaling behavior exhibited by the measure $\mu$. Unfortunately, accurately inferring the multifractal spectrum by reliably estimating the H\"{o}lder exponent point by point from real-world data and then computing the fractal dimension of each resulting iso-scaling set $E_\alpha$ is fraught with problems to the point of being impractical~\cite{salat2017multifractal} if not meaningless~\cite{evertsz1992multifractal}.

However, there exists an alternative description that is statistical rather than geometrical in nature and also practical. It is known as the \textbf{method of moments} and is concerned with inferring the scaling behavior of higher-order sample moments of $\mu$. In practice, this method involves dividing $E$ into increasingly more fine-grained partitions $E_r$ with equal-sized cells of size $r$ (\ie multiresolution decomposition), deriving for each partition $E_r$ and for a range of $q$-values the \textbf{partition functions} $Z(r,q)$ by computing the $q$th sample moments of $\mu$ defined on $E_r$, and then determining the so-called \textbf{structure function} $\tau(q)$ that describes the scaling behavior of these partition functions as the cell size $r$ tends to 0; that is, satisfies 
\begin{equation} 
Z(r,q) \approx r^{\tau(q)} ~~~\text{as}~~ r \to 0. 
\end{equation}

The two descriptions are related thanks to the so-called {\bf multifractal formalism}~\cite{olsen1995multifractal,riedi1995improved,salat2017multifractal} that asserts that under certain technical conditions, the multifractal spectrum $f(\alpha)$ is the {\em Legendre transform} of the structure-function $\tau(q)$ (and vice versa); that is, 
    $f(\alpha) = \inf_{q} (q\alpha - \tau(q))$. 
For example, this relationship is very useful when all that is needed from a multifractal analysis of real-world data is plausible evidence of whether or not the data is consistent with multifractal scaling behavior.
Indeed, by virtue of being the Legendre transforms of one another, a nonlinear structure function $\tau(q)$ implies a non-degenerate multifractal spectrum $f(\alpha)$ that is defined for some continuous range of $\alpha$-values on the real line (because each $q$ yields a value of $\alpha = \frac{\partial}{\partial q}\tau(q)$).
Thus, to obtain plausible evidence of $\mu$ being multifractal or monofractal, it suffices to check whether the structure function $\tau(q)$ is not linear (evidence for multifractal scaling, $\alpha$ taking a range of values) or linear (evidence for monofractal scaling, $\alpha$ taking a single value).

For structure functions $\tau(q)$ that are not linear, three values of particular interest, commonly referred to as {\bf generalized dimensions}, are often used to provide a coarse-grained characterization of multifractals. These values are (i) $D_0 := -\tau(0)$ which is the usual {\bf box-counting (or fractal) dimension} and provides information about how the data points or observations fill the physical space they occupy; (ii) $D_1 := \tau'(1^-)$ which is commonly referred to as the {\bf information dimension}, relates to Shannon’s entropy and captures how evenly the data points occupy the space, with larger values of $D_1$ indicating a more uniform density of the points; and (iii) $D_2 := \tau(2)$ which is sometimes called the {\bf correlation dimension}, measures how scattered the data is, with larger values of $D_2$ associated with higher compactness or less scattering. It can be shown that $D_2 \leq D_1 \leq D_0$, and one requirement for assessing the quality of estimators of these quantities is that the estimated quantities satisfy the same ordering relationship. 

\subsection{Conservative or semi-random cascades}
\label{ssec:cascade}

Mathematical modeling of observed multifractal scaling behavior in various real-world datasets was pioneered by B. Mandelbrot who first introduced a class of prescriptive or evocative models known as \textbf{multiplicative cascades}~\cite{mandelbrot:1990}. Following~\cite{ossiander2000statistical,gilbert-etal:2003}, a multiplicative cascade is an iterative process that fragments a given set into smaller and smaller pieces according to some rule and, at the same time, distributes the total ``mass'' associated with the given set according to another rule. The cascade's \textbf{generator} determines the redistribution of the set’s total mass at every iteration and can be either deterministic (as, for example, in the ``Cantor dust" model considered in~\cite{kohler2002observed}) or random as in the case of the multiplicative, multiscale innovation model (\eg see~\cite{nowak:1998,gilbert-etal99,barford2006toward}).  

Multiplicative cascades with the property that the generator preserves the total mass of the initial set at each stage of the construction are called \textbf{conservative cascades} or \textbf{semi-random cascades} and are of particular interest because
{\em (i)} their construction is closely connected to physical concepts such as splitting of a fixed mass (\eg of IP addresses) between regions (\eg address prefixes) and
{\em (ii)} a rich body of mathematical results and statistical theory has emerged over the years yielding powerful estimation techniques for naturally occurring phenomena based on conservative cascades~\cite{holley1992multifractal,ossiander2000statistical,resnick2003wavelet}.
For example, conservative cascades with generators $W$ that satisfy certain technical conditions have been shown theoretically to generate mathematical multifractals that can be characterized in terms of the cascade's generator $W$ (\ie $\tau(q)$ is a function of $W$), which in turn can be robustly inferred from real-world data. 

More precisely, consider the case of a fixed generator given by the random variable $W$ that has mean 1/2, is defined on the interval $(0,1]$, and is symmetric around the mean. Then, using results in~\cite{holley1992multifractal,ossiander2000statistical,gilbert-etal:2003}, for $q$-values below a critical value $q^{\star}$, the structure function $\tau (q)$ of the multifractal measure generated by the conservative cascade with generator $W$ and defined as the scaling exponent of the partition functions $Z(r,q)$ (\ie $Z(r,q) \sim r^{-\tau(q)}$ as $r \to 0$, see \S~2.2) is given by
\begin{equation}
\tau(q) = \log_2 E[W^q] + 1
\label{structure-fct}
\end{equation}
The relationship shows that in this case, the structure function $\tau(q)$ can be estimated via the generator $W$, meaning that the distribution of $W$ encodes the full spectrum of scaling behaviors exhibited by a single realization of the cascade. Importantly, for performing inference on the distribution of the cascade generator $W$ from real-world data that was presumably generated by this conservative cascade in the first place, we can leverage an existing statistical theory for different estimators~\cite{ossiander2000statistical,gilbert-etal:2003}. This theory provides strong consistency and central limit theorem results and allows, for example, for the computation of confidence intervals for the considered statistical estimators. 

Note that although the use of conservative cascades for describing the {\em temporal} nature of Internet traffic has been extensively explored in~\cite{feldmann-sigcomm98,riedi-lv:1997,mannersalo-norros:1997,riedi-etal99} (and subsequently largely refuted in~\cite{veitch-etal05}), their potential to model {\em spatial} aspects of addresses observed in Internet traffic, which is the focus of our work, has received less attention.

\section{Appendix: Repository of recent traffic traces}
\label{appx:traceRepo}

Our efforts to assemble our own collection of measured Internet traffic traces are driven by three requirements. First, to ensure the relevance of our work for today’s Internet, we require our traces to reflect ``modern” Internet traffic (\eg recorded within the last ten years). Second, following~\cite{paxson-floyd:ton2001} and using the term ``traffic invariant” to mean some facet of behavior of traffic which has been empirically shown to hold in a very wide range of network settings or environments, to examine whether or not an empirical property such as the multifractal structure of observed IP addresses in measured network traffic can be viewed as such an ``invariant”, we require the different traces to be collected from different network locations (e.g., backbone links vs. access links) and over periods of time that span a few years. Lastly, to further examine such invariants and their presence/absence in particular portions of network traffic, we also want our collection to include traces that differ by type of traffic (\eg Darknet vs generic). 

To this end, we amassed a collection of publicly available datasets, 
alongside a number of privately collected datasets recorded from different university campus border switches (\eg University of Oregon, UCSB) and other datasets that were provided by a third party (\eg Darknet traffic). These datasets are shown in Table~\ref{table:list_of_datasets} where the different traces are grouped into backbone-link traffic traces, access-link traffic traces, and Darknet traffic traces.
In particular, CAIDA refers to CAIDA’s uni-directional traffic traces which were recorded from an Internet backbone link in 2019~\cite{caida2019} and which are anonymized using CryptoPan prefix-preserving anonymization~\cite{cryptopan}; MAWI refers to (bi-directional) traffic traces from the MAWILab traffic anomaly archive \cite{mawilab} that were recorded at the border between the WIDE project and its parent ISP, span a period of six years (2015-2021) and are anonymized with tcpdpriv~\cite{tcpdpriv}; and
UO (UCSB) refers to captures of network traffic at the edge of the University of Oregon's (University of California Santa Barbara’s) campus network (data from both campus networks were anonymized prior to analysis using the prefix-preserving anonymization method CryptoPan~\cite{cryptopan}, thus ensuring that running our analysis of the observed IP addresses  before and after anonymization yields identical results).

\begin{center}
\begin{table*}
\centering
\begin{tabular}{ | c | c | c | c | c | c | }
\hline
 ToI/ToI Instances & Date & Duration & Type & IP Count & Packet Count \\ 
 \hline
 \hline
 \textbf{ToI: Backbone-link} \\ 
 \textbf{(uni-directional)} \\
 \hline
 \textit{CAIDA-dir-A} & 2019-01-17 5:00AM & 900s & pcap & 2,381,306 & 566M \\  
 \textit{CAIDA-dir-B} & 2019-01-17 5:00AM & 900s & pcap & 3,863,987 & 1.23B \\
 \hline
 \textbf{ToI: Backbone-link} \\ 
 \textbf{(bi-directional)} \\ 
 \hline 
 \textit{MAWI-20150202} & 2015-02-02 2:00PM & 900s & pcap & 5,571,625 & 99M \\
 \textit{MAWI-20150710} & 2015-07-10 2:00PM & 900s & pcap & 4,415,599 & 191M \\
 \textit{MAWI-20151002} & 2015-10-08 2:00PM & 900s & pcap & 4,818,370 & 135M \\
 \textit{MAWI-20180316} & 2018-03-16 2:00PM & 900s & pcap & 4,567,614 & 69M \\
 \textit{MAWI-20180807} & 2018-08-07 2:00PM & 900s & pcap & 4,635,311 & 73M \\
 \textit{MAWI-20181107} & 2018-11-07 2:00PM & 900s & pcap & 4,677,191 & 80M \\
 \textit{MAWI-20190901} & 2019-09-01 2:00PM & 900s & pcap & 4,689,835 & 87M \\
 \textit{MAWI-20210110} & 2021-01-10 2:00PM & 900s & pcap & 265,794 & 52M \\
 \textit{MAWI-20210614} & 2021-06-14 2:00PM & 900s & pcap & 180,532 & 74M \\
 \textit{MAWI-20211212} & 2021-12-12 2:00PM & 900s & pcap & 149,572 & 55M \\
 \hline
 \textbf{ToI: Access-link (UCSB)} \\
 \textbf{(incoming)} \\
 \hline
 \textit{UCSB-20220428} & 2022-04-28 12:15 & 900s & pcap & 194,498 & 1.02B \\
 \textit{UCSB-20220921} & 2022-09-21 19:25 & 900s & pcap & 112,164 & 1.05B \\
 \textit{UCSB-20221205} & 2022-12-05 20:15 & 900s & pcap & 186,575 & 1.1B \\
 \hline
 \textbf{ToI: Access-link (UO)} \\
 \textbf{(incoming)} \\
 \hline
 \textit{UO-20181106} & 2018-11-16 00:00 & 1 day & netflow & 1,805,085 & 9B \\
 \textit{UO-20190829} & 2019-08-29 00:00 & 1 day & netflow & 1,497,311 & 12B\\
 \textit{UO-20200213} & 2020-02-13 00:00 & 1 day & netflow & 786,607 & 36B \\
 \textit{UO-20211106} & 2021-11-06 00:00 & 1 day & netflow & 1,288,775 & 26B\\
 \textit{UO-20220517} & 2022-05-17 00:00 & 1 day & netflow & 668,817 & 23B\\
\hline
\end{tabular}
\caption{Our collection of recent traffic traces.
}
\label{table:list_of_datasets}
\end{table*}
\end{center}

\pagebreak
\section{Appendix: Additional Empirical Evidence in Support of our Physical Explanation}
\label{sec:appdx-physical-explanation}

This section provides further evidence and details in support of our argument from \S~\ref{sec:physical-explanation} that the cascade model is an inherent part of the address allocation process.

\subsection{Historical global-level IPv4 allocation}
\label{ssec:appx-historicalv4}

\begin{figure}[!htb]
    \centering
    \includegraphics[width=0.5\columnwidth]{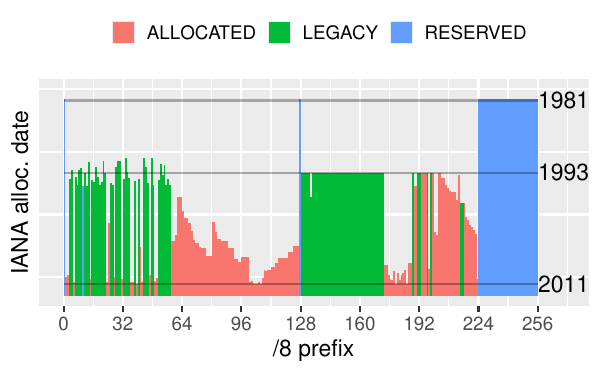}
    \caption{Global view of the IANA IPv4 Address Space Registry showing how different regions have been allocated by IANA over time.}
    \label{fig:iana_prefixAllocStatus}
\end{figure}

Figure~\ref{fig:iana_prefixAllocStatus} shows for each /8 block (x-axis) the year in which is was allocated by the IANA (y-axis) as well as the allocation type (color).
The majority of top-level allocations fall into three distinct historical phases which roughly correspond to their allocation type.

In the first phase, several well-known ``reserved'' regions were allocated during standardization of IPv4 in 1981 (\eg local identification in 0.0.0.0/8, multicast in 224.0.0.0/4)~\cite{rfc790, rfc791}.
These regions typically serve intra-organization protocols and are not routed on the public Internet.
In real Internet traces, theses regions are typically entirely empty corresponding to being assigned zero, or near-zero mass in the MCC model.

In the second phase, during the early days of the Internet, addresses were divided into three distinct classes and IANA allocated addresses from different classes directly to organizations operating network infrastructure.
As described in RFC~790~\cite{rfc790}, the 32 bits of the IPv4 address were divided into a ``network'' portion and a ``local address'' portion allowing for a small number of large-size ``class A'' networks (with 8 network bits, 24 local address bits), a medium number of medium-size ``class B'' networks (with 16 network bits, 16 local bits), and a large number of small-size ``class C'' networks (with 24 network bits, 8 local bits).
More importantly, different regions of the global space were blocked off for each class of network: all class A networks were allocated in 0.0.0.0/1; all class B networks in 128.0.0.0/2, and all class C networks in 192.0.0.0/3.\footnote{For easier comparison we use the modern CIDR notation of these regions instead of the bit-level prefixes described in RFC~790 (note that CIDR was not yet invented when RFC~790 was written in 1981).}
Finally, addresses in each class were allocated based on demand in more-or-less sequential order.
RFC~790 includes an snapshot showing that in 1981 the first 44 class A networks were already allocated and the remaining 83 were ``Unassigned''.
Figure~\ref{fig:iana_prefixAllocStatus} illustrates the impact of these early allocation policies on current address space allocation by showing a clear clustering of ``Legacy'' networks extending upward from the first class A and class B networks.

Finally, in the third phase, allocation policies were fundamentally changed by two key developments in 1993.
The first development was the establishment of Regional Internet Registries (RIRs) as described in RFC~1466~\cite{rfc1466}---instead of acting as the sole arbiter of addresses, IANA would allocate /8 blocks to RIRs which would then manage allocation of addresses within these blocks to organizations within their geographic regions.
Later that same year RFC~1519~\cite{rfc1519} proposed an approach to address space organization called Classless Inter-Domain Routing (CIDR)---instead of allocating fixed-size networks in certain regions, CIDR allows variable numbers of network bits across the address space and annotates network addresses with their ``prefix length'' indicating how many bits form the network portion (\eg 128.0.0.0/8 indicates a network portion of 8-bits with the decimal value 128).
On one hand, the development of RIRs added a step in the cascade process determining how IPv4 addresses are allocated.
On the other hand, the development and subsequent popularization of CIDR enabled the modern notion of prefix-level containment as a way to hierarchically structure relationships along the address allocation cascade.
As shown in Figure~\ref{fig:iana_prefixAllocStatus}, after 1993 IANA pursued distinct sequential strategies with their top-level allocations until the last block was allocated in 2011~\cite{ipv4exhaustion}.
Moreover, these sequentially-allocated regions correspond directly to single-RIR clusters shown in Figure~\ref{fig:iana_maxAgg}, \eg ARIN's cluster around 64.0.0.0/5 appears to have been allocated from IANA between 1999 and 2004 at a rate of roughly 2 /8-blocks per year.
Overall, IANA followed strong sequential patterns when allocating blocks to RIRs and potentially pre-allocated contiguous clusters of blocks for RIRs with higher expected growth like RIPE, APNIC, and ARIN.

\subsection{Temporal stability of the RIR-level prefix-inclusion tree}
\label{ssec:appx-rir-level}

\begin{figure}[!htb]
    \vspace{-0.3cm}
    \centering
    \includegraphics[width=0.45\linewidth]{fig/global_whois_rir_legend.pdf}
    
    \begin{subfigure}{0.3\linewidth}
        \includegraphics[width=0.9\columnwidth]{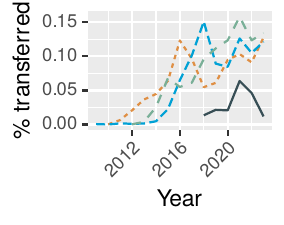}
        \vspace{-0.4cm}
        \caption{IPv4}
        \vspace{-0.2cm}
    \end{subfigure}
    \begin{subfigure}{0.3\linewidth}
        \includegraphics[width=0.9\columnwidth]{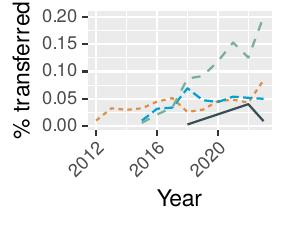}
        \vspace{-0.4cm}
        \caption{IPv6}
        \vspace{-0.2cm}
    \end{subfigure}
    \caption{Number of transfers per year normalized by each RIR's total number of IPv4 or IPv6 network records in 2024.}
    \label{fig:transfers:total}
    \vspace{-0.3cm}
\end{figure}

Next, we investigate how prefix transfers could impact the structure of the RIR-level prefix-inclusion tree using separate public datasets documenting all prefix transfers (\eg~\cite{arinTransfers}).
Figure~\ref{fig:transfers:total} shows the total number of transfers registered each year since 2009 by each RIR normalized by the total number of network records in 2024.\footnote{We do not differentiate between intra- and inter-RIR transfers because they do not have any fundamental difference w.r.t. structure of the prefix-inclusion tree.}
For most RIRs the normalized rate of transfers gradually increases till reaching a plateau around 0.1\% per year in 2017 for IPv4 and 0.05\% per year in 2018 for IPv6.
Overall, these results indicate that although transfers are a persistent aspect of address space allocation process, their net effect on address space allocations is relatively small (less that 2\% per year in all cases).

We also note two interesting peculiarities visible in Figure~\ref{fig:transfers:total}.
First, RIPE has an unusually large number of IPv6 transfers even after normalization.
This is likely due to RIPE's relatively loose IPv6 transfer policy geared towards driving IPv6 adoption which was recently flagged by the RIPE community as potential ``stockpiling''~\cite{ripeIPv6Stockpiling} and subsequently induced updated policies starting in 2024~\cite{ripeIPv6NewPolicy}. 
Second, AFRINIC sees extremely low transfer volume compared to other RIRs even after normalizing by its relatively smaller share of total network records.
This is likely a confluence of several factors including AFRINIC's lack of an inter-RIR transfer policy (despite multiple efforts~\cite{afrinicInterProp}), relatively strict intra-RIR transfer policy (\eg for IPv4 the transfer recipient must justify need for the address space), and continued possession of unused IPv4 ranges allocated from IANA~\cite{afrinicTransferPolicy,afrinicTransferTrading}.
These peculiarities hint at the complexity of address space transfers and illustrate various limiting forces on transfer policy.

\begin{figure}[!htb]
    \centering
    \begin{subfigure}{0.3\textwidth}
        \includegraphics[width=0.9\columnwidth]{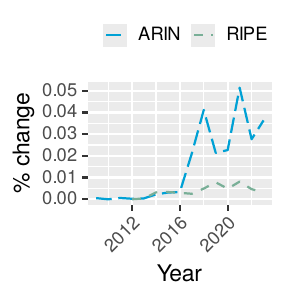}
        \vspace{-0.4cm}
        \caption{IPv4}
        \vspace{-0.2cm}
    \end{subfigure}
    \begin{subfigure}{0.3\textwidth}
        \includegraphics[width=0.9\columnwidth]{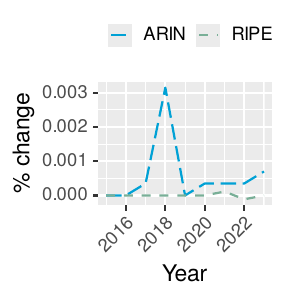}
        \vspace{-0.4cm}
        \caption{IPv6}
        \vspace{-0.2cm}
    \end{subfigure}
    \caption{Change in total number of prefixes relative to 2024.}
    \label{fig:transfers:relative}
    \vspace{-0.3cm}
\end{figure}

In addition to changes in ownership, transfers can also potentially change the structure of the prefix-inclusion tree by splitting parent prefixes into child prefixes and then assigning different organization to each child, or by merging adjacent children into a parent prefix assigned to a single organization.
In addition to the standard source and recipient organization information, ARIN and RIPE also report the original source address ranges that were consumed in each transfer (AFRINIC and APNIC, only report the final transferred address ranges making this analysis impossible). 
We quantify the impact of such changes on the overall tree structure by computing the total number of prefixes released from any organization and the total number of prefixes re-assigned to any organization each year.\footnote{Note that this metric over-estimates the amount of change since sometimes larger prefixes are transferred incrementally in blocks over year boundaries (\eg the transfer of 3.0.0.0/8 from GEC to Amazon in late 2017 and early 2018).}
Figure~\ref{fig:transfers:relative} shows the difference between the number of released and the number of re-assigned prefixes each year normalized by the total number of presently registered prefixes (in 2014). 
For almost every year, the net number of prefixes assigned across all organizations increases as a result of transfers due to splitting of larger parents into smaller children.
However, given the large number of total allocations (\eg for ARIN $\sim$3.2M IPv4, $\sim$290k IPv6 prefixes in 2024), the total change in structure introduced by these transfers remains quite small.
Interestingly, transfers in ARIN produce markedly more structural change compared to transfers in RIPE.
We also note that the particular spike in IPv6 transfers in ARIN observed in 2018 is due to two particular corporate re-organizations which split large undifferentiated address ranges into large numbers of smaller specific-use ranges.\footnote{Manually inferred based on similarities of organization name. For example the domain-name registry ``Afilias Canada, Corp.'' transfered prefixes to ``Afilias, Inc.'' as part of internal reorganization likely as part of its move to supply registration for the .au TLD in 2018~\cite{afiliasWiki}.}
Overall, these observations imply that even though prefixes are regularly transferred between organizations, the overall structural relationships (including which prefixes and granularities are allocated) are relatively constant.

\section{Appendix: Details of the Conservative Cascade as a Generative Model for IP Addresses}

This section provides additional details and supporting evidence for our proposed method for generating sets of IP addresses based on a conservative cascade process introduced in \S~\ref{sec:generative}.
Some text is repeated to retain a complete discussion.

\subsection{Impact of finite discreteness on fitting $W$}
\label{ssec:appx-fittingDetails}

\begin{figure}[!htb]
    \centering
    \includegraphics[width=0.9\linewidth]{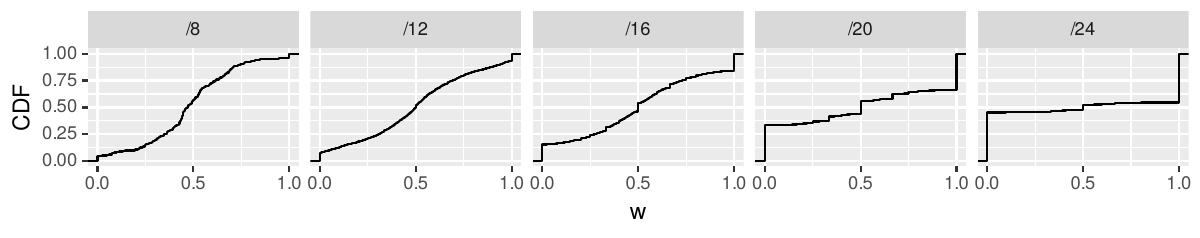}
    \caption{Distribution of the raw values of $w_{l,j}$ for the CAIDA500k dataset at various prefix lengths.}
    \label{fig:weigthsCaidaRaw}
\end{figure}

To illustrate the process of directly fitting a distribution for $W$ based on observed values $w_{l,j}$, we compute the values $w_{l,j}$ for the real-world CAIDA500k dataset and show their distribution for five example prefix lengths (in particular $l = 8, 12, 16, 20, 24$) in Figure~\ref{fig:weigthsCaidaRaw}.
These distributions are almost exactly symmetric (\ie mean 0.5) and closely resemble well-known distributions that take values in $[0,1)$ such as the Logit-normal distribution or Beta distribution, especially at short to medium prefix lengths (\eg /8 - /16).
Based on this finding and for the purpose of concreteness, we consider in this paper symmetric Logit-normal distributions due to their single, easily-estimated parameter, though our methods could easily be applied to other $[0,1)$ distributions such as symmetric Beta distributions.

Despite suggesting a general distributional shape, Figure~\ref{fig:weigthsCaidaRaw} also exposes several salient challenges in fitting the $w_{i,j}$ to continuous parametric models like the Logit-normal distribution.
First, because of the increased sparseness of the common ancestor tree when moving to longer prefix lengths (\eg $l=24$ and longer), a majority of longer prefixes have only a single observed address and hence the only possible values for $w_{i,j}$ are 0 and 1 leading to an increasingly pronounced modes (\ie step-function shape) at 0 and 1.
Second, even when they contain more than a single address, longer prefixes are still limited to a relatively small number of distinct addresses (\eg a maximum of 256 addresses in a /24 prefixes), resulting in less pronounced but clearly recognizable modes at 0.5, 0.25, 0.75, and so forth.
Finally, since there are relatively few prefixes at shorter prefix lengths (\eg 16 /4 IPv4 prefixes), the distribution becomes increasingly noisy at smaller prefix lengths due to the small number of samples.

To address these challenges, we pre-process the $w_{l,j}$'s with the following filtering and transformations before performing the standard Logit-normal fit.
First, we remove all $w_{i,j}$'s that correspond to prefixes with only a single address.
From the conservative cascade perspective, these prefixes have reached a terminal point and have no further capacity for diving of mass, hence the derived $w_{i,j}$'s do not provide any information about how mass is divided by $W$.
Second, we replace any remaining instances where $w_{i,j}$ is 0 with 1 / $2n$ and where $w_{i,j}$ is 1 with 1 - 1 / $2n$ where $n$ is the number of addresses being divided by $w_{i,j}$.
On the one hand, when considering a generator $W$ with a continuous distribution, we have $P[W=0]=P[W=1]=0$. At the same time, the values 0 and 1  %
produce undefined values in the transformation required by Logit-normal fitting.\footnote{In particular, if $Z$ has a symmetric Logit-normal distribution, then $\ln(Z / (1 - Z))$ has a mean-zero Normal distribution.} For these reasons, we replace both of these values with the nearest fractional value that rounds to 0 or 1.
Finally, we only use values of $w_{l,j}$ from $8 \leq l \leq 16$ (\ie only prefixes between /8 and /16) to avoid the noisy behavior at shorter prefix lengths and the more pronounced perturbations by discrete values at longer prefix lengths.

\begin{figure}[!htb]
    \vspace{-0.2cm}
    \centering
    \includegraphics[width=0.9\linewidth]{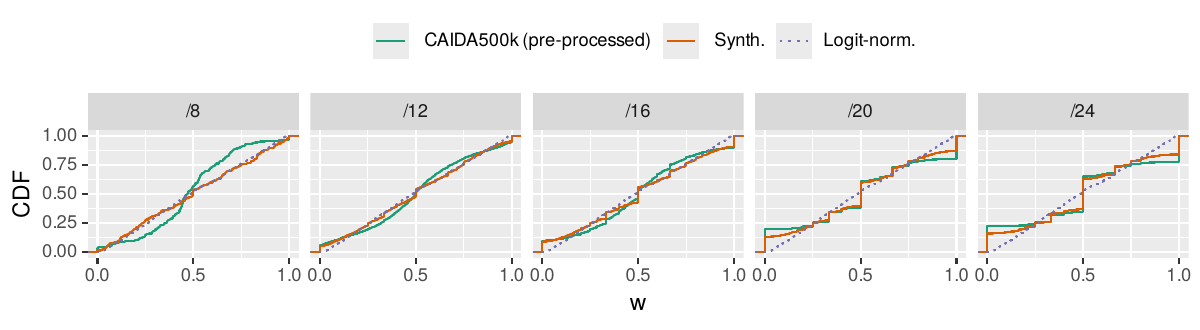}
    \vspace{-0.5cm}
    \caption{Fitted Logit-normal distribution for the pre-processed $w_{i,j}$'s for the CAIDA500k dataset, along with a numerically computed Logit-normal distribution and a fitted Logit-normal distribution for the pre-processed $w'_{i,j}$'s for a dataset synthesized with the method of \S~\ref{ssec:generating}.}
    \label{fig:weigthsFit}
\end{figure}

By applying these pre-processing steps and the usual standard-deviation estimator to the Normal-transformed $w_{i,j}$'s, we find that the CAIDA500k data's generator $W$ is well-modeled by a Logit-normal with parameter $\sigma = 1.61$.
Figure~\ref{fig:weigthsFit} shows the distribution of the pre-processed version $w_{i,j}$ for the  CAIDA500k data (solid red), a numerically computed Logit-normal distribution with $\sigma = 1.61$ (dotted green), and a synthetic finite discrete conservative cascade as described in the next section (dashed blue).
We observe that the numerically computed Logit-normal distribution is close to the observed $w_{i,j}$ for longer prefix lengths (\eg /12 and longer), but has higher variance compared to the fitted generator for the CAIDA500k datasets for shorter prefix lengths (\eg /8).
This is likely caused by the fact that there are relatively few observations of $w_{i,j}$ for shorter prefix lengths which biases the fit towards longer prefix lengths.
We also note that after pre-processing, the observed data appears to have slightly lower variance at shorter prefix lengths, possibly due to less sparsity in the common ancestor tree.
Due to limited space, we leave a deeper investigation of these issues (\eg weighting the fit and/or exploring models where the generator's variance can change as a function of prefix length) to future work.

\subsection{Pictorial evidence for multifractal structure of synthetic IP addresses}
\label{ssec:appx-pictorialSynthEvidence}

\begin{figure}[htb!]
    \centering
    \begin{subfigure}{0.47\linewidth}
        \centering
        \includegraphics[width=0.9\columnwidth]{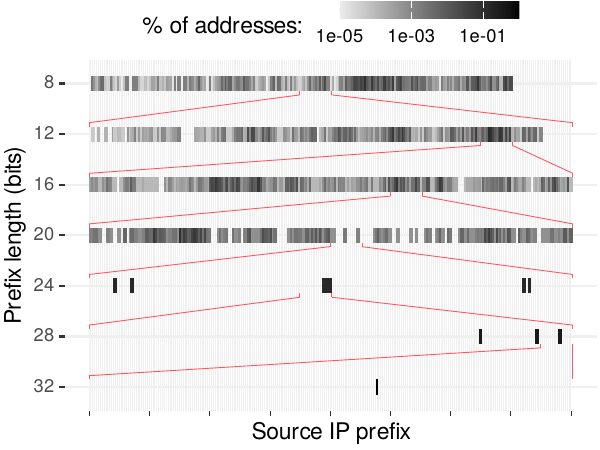} %
        \caption{Path to 125.168.127.152.}
    \end{subfigure}~
    \begin{subfigure}{0.47\linewidth}
        \centering
        \includegraphics[width=0.9\columnwidth]{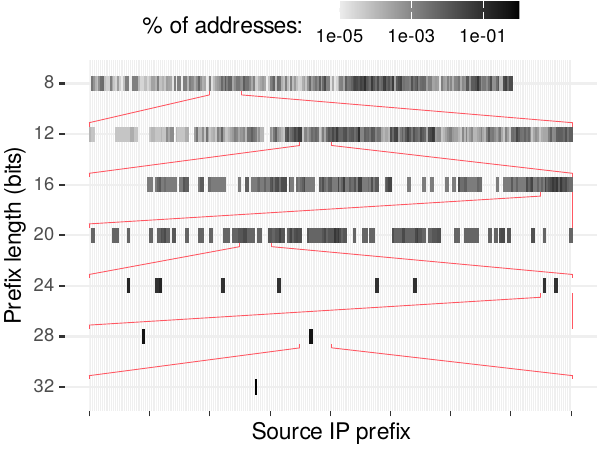}
        \caption{Path to 71.245.247.88}
    \end{subfigure}
    \vspace{-0.4cm}
    \caption{1D Pictorial evidence of multifractal scaling in a synthetic finite discrete conservative cascade with logit-normal generator.} 
    \label{fig:generaticPicture}
\end{figure}

Finally, as an initial qualitative exploration of the structural properties of our synthetically generated  addresses, we show in Figure~\ref{fig:generaticPicture} similar pictorial evidence for multifractal scaling as presented for the CAIDA500k dataset in Figure~\ref{1Dfigure:compare}.
These examples demonstrate that our synthetically generated addresses exhibit similar cluster-within-cluster structure as observed in CAIDA500k at multiple prefix levels and multiple regions of the address space.
Moreover, notable reserved regions (like the multicast address space visible in the top right of each figure) are also empty as in the CAIDA500k data.

\section{Appendix: Extended details of our analysis and results}

\subsection{Details of impact of finite discreteness on method-of-moments analysis of IP addresses}
\label{appx:analysisDetails}

\begin{figure}[!htb]
    \vspace{-0.3cm}
    \centering
    \includegraphics[width=0.65\linewidth]{fig/Zs_legend.pdf}
    
    \begin{subfigure}{0.25\textwidth}
        \includegraphics[width=0.9\columnwidth]{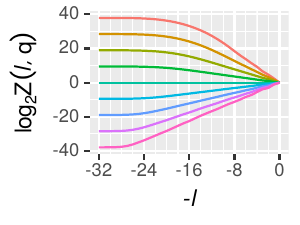}
        \caption{$\sigma = 1.0$}
    \end{subfigure}~
    \begin{subfigure}{0.25\textwidth}
        \includegraphics[width=0.9\columnwidth]{fig/Zs_LogitNormReserved1.61.pdf}
        \caption{$\sigma = 1.61$}
    \end{subfigure}~
    \begin{subfigure}{0.25\textwidth}
        \includegraphics[width=0.9\columnwidth]{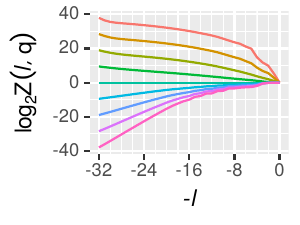}
        \caption{$\sigma = 3.16$}
    \end{subfigure}
    \caption{``Partition function'' plots for finite discrete conservative cascades based on logit-normal generators with different variance parameter $\sigma$.}
    \label{fig:partitionFunctionsSynths}
\end{figure}

To understand the root causes of the presence of different, nonlinear behaviors in the partition functions of Figure~\ref{fig:partitionFunctions}, we leverage the increased visibility into conservative cascade models afforded by our synthetic generation method.
Our hypothesis is that these nonlinear behaviors are artifacts of ``edge-cases" that arise as a result of the finite and discrete nature of the IP space. While such spaces differ in theory from the continuous settings that are commonly considered in the multifractal analysis literature and allow for a rigorous treatment of mathematical limits and asymptotic behaviors, in practice, interpreting them as discrete approximations of suitably selected continuous counterparts has the potential of utilizing the same tools and methods that have been developed for the continuous case, albeit modified for or tailored to the discrete case.   
To explore this hypothesis, we use our finite discrete generation method to synthesize additional conservative cascades with generators that have a variance parameter $\sigma$ below (1.0) and above (3.16) the ``best-fit'' value for CAIDA500k estimated at 1.61 (see \S~\ref{sec:generative}) and examine how selecting this free parameter of our cascade model impacts the fundamental scaling behavior as revealed by the computed partition functions.
Figure~\ref{fig:partitionFunctionsSynths} compares the partition functions of all three of these synthetic cascades showing how the lower value of $\sigma$ tends towards the uniform distribution (\eg as shown in Figure~\ref{fig:partitionFunctions:uniform}) while the higher value of $\sigma$ increase the apparent non-linearity.

\begin{figure}[!htb]
    \vspace{-0.3cm}
    \centering
    \begin{subfigure}{0.25\textwidth}
        \includegraphics[width=0.9\columnwidth]{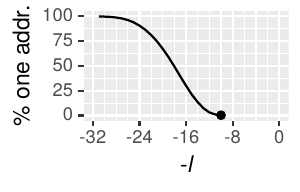}
        \caption{$\sigma = 1.0$}
    \end{subfigure}~
    \begin{subfigure}{0.25\textwidth}
        \includegraphics[width=0.9\columnwidth]{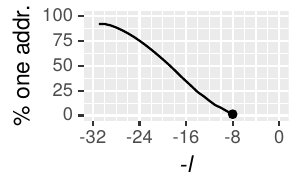}
        \caption{$\sigma = 1.61$}
    \end{subfigure}~
    \begin{subfigure}{0.25\textwidth}
        \includegraphics[width=0.9\columnwidth]{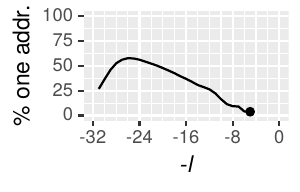}
        \caption{$\sigma = 3.16$}
    \end{subfigure}
    \caption{Fraction of how many prefixes at each level have only one address for the same synthetic conservative cascades shown in Figure~\ref{fig:partitionFunctionsSynths}}
    \label{fig:singularFracs}
\end{figure}

\begin{figure}[!htb]
    \vspace{-0.3cm}
    \centering
    \begin{subfigure}{0.25\textwidth}
        \includegraphics[width=0.9\columnwidth]{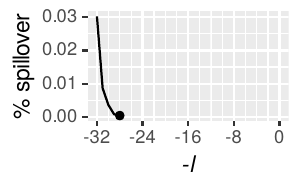}
        \caption{$\sigma = 1.0$}
    \end{subfigure}~
    \begin{subfigure}{0.25\textwidth}
        \includegraphics[width=0.9\columnwidth]{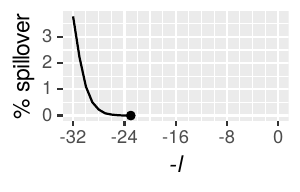}
        \caption{$\sigma = 1.61$}
    \end{subfigure}~
    \begin{subfigure}{0.25\textwidth}
        \includegraphics[width=0.9\columnwidth]{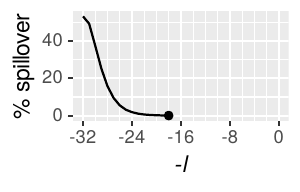}
        \caption{$\sigma = 3.16$}
    \end{subfigure}
    \caption{Fraction of how many prefixes were ``spilled over'' in the synthetic conservative cascades shown in Figure~\ref{fig:partitionFunctionsSynths}}
    \label{fig:spillovers}
\end{figure}

First, we consider the change in appearance (or nonlinearity) observed at shorter prefix lengths.
This change appears to induce the strongest nonlinearity in the smaller negative moments which are more sensitive to prefixes with fewer addresses.
Accordingly, we consider prefixes with only a single address.
Intuitively, these prefixes introduce a markedly different type of scaling behavior: in fact they {\em do not} scale but retain exactly the same mass up to termination of the cascade.
Figure~\ref{fig:singularFracs} shows the fraction of all prefixes that have only one address at each prefix length with a black dot marking the shortest prefix length where this fraction is first non-zero.
This plot confirms that prefixes with a single address are responsible for the change in scaling behavior at short prefix lengths because the point where this fraction is first non-zero corresponds exactly to the short-length change in scaling behavior shown in the partition functions (\eg for $\sigma = 1.0$, the first single-address prefixes emerge at /9 whereas for $\sigma = 3.16$ the first single-address prefixes emerge at /5, both of which correspond to the respective nonlinearities shown at shorter prefix lengths in Figure~\ref{fig:partitionFunctionsSynths}).

Next, we investigate the change in scaling observed at longer prefix lengths.
This change appears to induce the strongest nonlinearity in the larger positive moments which are more sensitive to prefixes with larger numbers of addresses.
Intuitively, the spilling over of addresses when the generator $W$ attempts to assign too much address ``mass'' to a (smaller) longer-length prefix could induce different scaling among larger prefixes.
Figure~\ref{fig:spillovers} shows the fraction of all prefixes that have one or more addresses spilled over to their child in our finite discrete conservative cascade at each prefix length longer than /16, again with a marker drawn where this fraction is first non-zero.\footnote{We focus here on longer-length prefixes and omit spillover caused at shorter prefix lengths by large reserved address blocks because the latter do not appear to impact scaling.}
This plot confirms that the inherently limited capacity of smaller prefixes at longer prefix lengths likely induces the change in scaling behavior shown in the left tail end in the partition functions (\eg for $\sigma = 1.0$, the first spillover occurrence is at /28 whereas for $\sigma = 3.16$ the first spillover occurrence is at /18, both of which correspond almost exactly to the nonlinearities shown at longer prefix lengths in Figure~\ref{fig:partitionFunctionsSynths}).

\subsection{Summary results over all datasets}
\label{appx:resultsSummary}

Table~\ref{table:analysis-results} shows the results of applying our adaptation of the method of moments (see \S~\ref{sec:evidence}) to estimate the three generalized dimensions for each trace in our repository. These quantities refine the binary decision ``multifractal: yes or no" that results from an application of the method of moments and produces a "yes" for all the listed traces. Despite some variation in these quantities across the different traces, they consistently provide evidence for multifractal scaling (\eg all dimensions less than one), in stark contrast to their counterparts for the synthetic monofractal trace UNIFORM500k (see Table~\ref{table:gen-dim} in \S~\ref{ssec:multifractalAsNewInvar}). As in Table~\ref{table:gen-dim}, standard deviation is shown for each $\tilde{\tau}(q)$. For $D_1$ we provide only a single value because it is estimated through least-squares and does not have a well-established notion of confidence. In particular, note that in agreement with the theory (see \S~\ref{ssec:multifracatal-analysis}), the estimated generalized dimensions $D_0, D_1$, and $D_2$ for all but one of the datasets satisfy the ordering relationship $D_0 \leq D_1 \leq D_2$, with dataset CAIDA-dir-B being the exception.

\begin{center}
\begin{table*}
\centering
\begin{tabular}{ | c | c | c | c | c | }
\hline
ToI & $D_0$ & $D_1$ & $D_2$ & $\tilde{\tau}(1) = 0?$ \\ 
Instances & (fractal dim.) & (information dim.) & (correlation dim.) &  \\
 \hline
 \hline 
 \textit{CAIDA-dir-A} & 0.92$\pm$0.001 & 0.81 & 0.68$\pm$0.009 & \checkmark \\  
 \textit{CAIDA-dir-B} & 0.62$\pm$0.004 & 0.62 & 0.66$\pm$0.015 & \checkmark \\
 \hline
 \hline 
 \textit{MAWI-20150202} & 0.94$\pm$0.001 & 0.88 & 0.80$\pm$0.021 & \checkmark \\
 \textit{MAWI-20150710} & 0.94$\pm$0.001 & 0.89 & 0.70$\pm$0.031 & \checkmark \\
 \textit{MAWI-20151008} & 0.94$\pm$0.001 & 0.89 & 0.74$\pm$0.030 & \checkmark \\
 \textit{MAWI-20180316} & 0.90$\pm$0.002 & 0.88 & 0.85$\pm$0.005 & \checkmark \\
 \textit{MAWI-20180807} & 0.91$\pm$0.001 & 0.89 & 0.88$\pm$0.003 & \checkmark \\
 \textit{MAWI-20181107} & 0.91$\pm$0.002 & 0.88 & 0.88$\pm$0.003 & \checkmark \\
 \textit{MAWI-20190901} & 0.91$\pm$0.001 & 0.88 & 0.87$\pm$0.003 & \checkmark \\ 
 \textit{MAWI-20210110} & 0.87$\pm$0.002 & 0.71 & 0.51$\pm$0.025 & \checkmark \\
 \textit{MAWI-20210614} & 0.86$\pm$0.002 & 0.68 & 0.36$\pm$0.021 & \checkmark \\
 \textit{MAWI-20211212} & 0.85$\pm$0.002 & 0.64 & 0.30$\pm$0.017 & \checkmark \\
 \hline
 \hline
 \textit{UCSB-20220428} & 0.87$\pm$0.002 & 0.62 & 0.15$\pm$0.012 & \checkmark \\
 \textit{UCSB-20220921} & 0.86$\pm$0.001 & 0.61 & 0.14$\pm$0.010 & \checkmark \\
 \textit{UCSB-20221205} & 0.86$\pm$0.001 & 0.61 & 0.16$\pm$0.013 & \checkmark \\
 \hline
 \hline
 \textit{UO-20181116} & 0.90$\pm$0.001 & 0.77 & 0.63$\pm$0.010 & \checkmark \\
 \textit{UO-20190829} & 0.93$\pm$0.001 & 0.79 & 0.63$\pm$0.012 & \checkmark \\
 \textit{UO-20200213} & 0.91$\pm$0.001 & 0.78 & 0.56$\pm$0.019 & \checkmark \\
 \textit{UO-20211106} & 0.92$\pm$0.001 & 0.78 & 0.52$\pm$0.025 & \checkmark \\
 \textit{UO-20220517} & 0.84$\pm$0.002 & 0.69 & 0.45$\pm$0.027 & \checkmark \\
\hline
\end{tabular}
\caption{Multifractal analysis results for all traffic traces listed in Table~\ref{table:list_of_datasets}.
}
\label{table:analysis-results}
\end{table*}
\end{center}

\begin{figure}[!htb]
    \vspace{-0.3cm}
    \centering
    \begin{subfigure}{0.25\textwidth}
        \includegraphics[width=0.9\columnwidth]{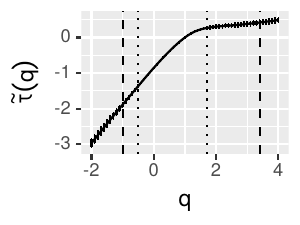}
        \caption{MAWI-20211212}
        \label{fig:partitionFunctionsAppx:mawi}
    \end{subfigure}~
    \begin{subfigure}{0.25\textwidth}
        \includegraphics[width=0.9\columnwidth]{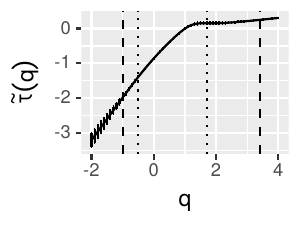}
        \caption{UCSB-20221205}
        \label{fig:partitionFunctionsAppx:ucsb}
    \end{subfigure}~
    \begin{subfigure}{0.25\textwidth}
        \includegraphics[width=0.9\columnwidth]{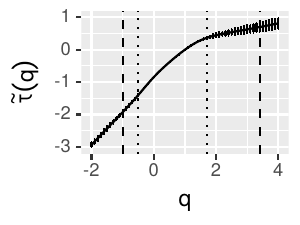}
        \caption{UO-20220517}
        \label{fig:partitionFunctionsAppx:uo}
    \end{subfigure}
    \caption{$\tilde{\tau}(q)$ (the estimated ``structure functions'') for the latest datasets from our three non-CAIDA data sources with 95\% confidence intervals. Dashed lines show the critical range for convergence, dotted lines show the critical range for normalicy.}
    \label{fig:structureFunctionsAppx}
\end{figure}

We also show in Figure~\ref{fig:structureFunctionsAppx} the $\tilde{\tau}(q)$ estimates for the latest samples from each of the three non-CAIDA data sources.
The clear non-linearity of these plots provides strong evidence for multifractal scaling in IP addresses observed at these different networks.
Moreover, the $\tilde{\tau}(q)$ estimates have different absolute shapes for each network providing a hint of how multifractal analysis can produce {\em quantitative} fingerprints of the distinct structure of IP addresses observed from different vantage points.

\pagebreak
\section{Appendix: Analysis of Specific Structural Changes Induced by Anomalous Addresses}
\label{sec:appx-anomalousChanges}

\begin{figure}[!htb]
    \centering
    \begin{subfigure}[t]{0.48\textwidth}
    \includegraphics[width=0.9\linewidth]{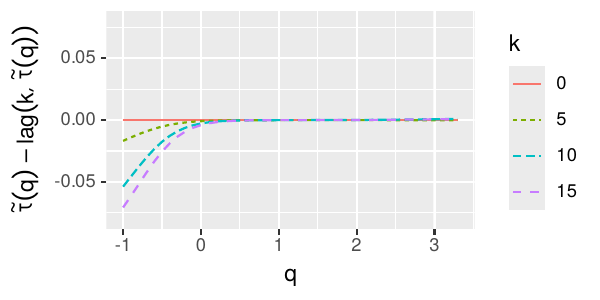}
    \caption{Anomalous addresses.}
    \label{fig:anomTauQs:anom}
    \end{subfigure}~
    \begin{subfigure}[t]{0.48\textwidth}
    \includegraphics[width=0.9\linewidth]{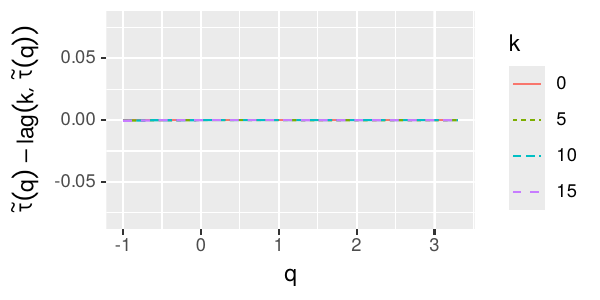}
    \caption{Control addresses.}
    \label{fig:anomTauQs:control}
    \end{subfigure}
    \caption{Comparison of initial and lagged versions of $\tilde{\tau}(q)$. The change induced by anomalous addresses shows up as a steepening of the negative moments for longer lags whereas the control addresses have no noticeable impact.}
    \label{fig:anomTauQs}
\end{figure}

To illustrate how the changes in anomalous score observed in \S~\ref{sec:implications} relate to changes in the underlying structure function estimated by $\tilde{\tau}(q)$, we show in Figure~\ref{fig:anomTauQs} the difference between the current estimate and its lagged version for several different lags (\ie values of $k$).
On the one hand, for ``control'' addresses (Figure~\ref{fig:anomTauQs:control}), we observe this difference is flat, indicated essentially no change in $\tilde{\tau}(q)$.
On the other hand, for the anomalous addresses (Figure~\ref{fig:anomTauQs:anom}), a distinct difference emerges for negative $q$-values in particular.
Given the fact that these negative $q$ capture the scaling behavior of prefixes with small numbers of addresses, this result is consistent with the idea that our approach is able to detect subtle changes in scaling structure introduced by only a few newly observed addresses.

\end{document}